%% file: paper.tex
\documentclass{article}
\usepackage[margin=1.25in]{geometry}

\input{header}

\author{Paul Friedrich$^{1,2}$, Barna Pásztor$^{1,2,3}$, Giorgia Ramponi$^{1,2}$}

\date{\centering\small $^1$ETH AI Center, Zurich, Switzerland \\
$^2$Department of Informatics, University of Zurich, Switzerland \\
$^3$Department of Computer Science, ETH Zurich, Switzerland}

\title{\bf Learning Collusion in Episodic, Inventory-Constrained Markets\Large}

\begin{document}

\maketitle 

\def\thefootnote{}\footnotetext{Code: \url{https://github.com/pfriedric/EpisodicCollusion}. Correspondence: Paul Friedrich, \href{mailto:paul.friedrich@uzh.ch}{paul.friedrich@uzh.ch}.}\def\thefootnote{\arabic{footnote}}

\begin{abstract}
\noindent Pricing algorithms have demonstrated the capability to learn tacit collusion that is largely unaddressed by current regulations. Their increasing use in markets, including oligopolistic industries with a history of collusion, calls for closer examination by competition authorities. In this paper, we extend the study of tacit collusion in learning algorithms from basic pricing games to more complex markets characterized by perishable goods with fixed supply and sell-by dates, such as airline tickets, perishables, and hotel rooms. We formalize collusion within this framework and introduce a metric based on price levels under both the competitive (Nash) equilibrium and collusive (monopolistic) optimum. Since no analytical expressions for these price levels exist, we propose an efficient computational approach to derive them. Through experiments, we demonstrate that deep reinforcement learning agents can learn to collude in this more complex domain. Additionally, we analyze the underlying mechanisms and structures of the collusive strategies these agents adopt.
\end{abstract}

\section{Introduction}
Algorithms are increasingly replacing humans in pricing decisions, offering improved revenue management and handling of complex dynamics in large-scale markets such as retail and airline ticketing. These algorithms, whether programmed or self-learning, can engage in tacit collusion charging \emph{supra-competitive} prices (i.e., above the competitive level) or limiting production without explicit agreements.
For example, algorithmic pricing in Germany led to a 38\% increase in fuel retailer margins after adoption~\citep{assadAlgorithmicPricingCompetition2024}.
Our study is primarily motivated by airline revenue management (ARM), a market with \$800 billion in annual revenue and thin profit margins.
Airlines have already been under regulatory scrutiny~\citep{EUreg2019} due to evidence of tacit collusion even before the introduction of algorithmic pricing~\citep{borenstein1994competition} but the current trend of moving towards algorithmic pricing~\citep{koenigsbergEasyJetAirlines2004, Razzaghi2022} could lead to further cases.

\looseness=-1
Tacit collusion is maintained without explicit communication or agreement between sellers, therefore it eludes detection and often falls outside the scope of current competition laws. These concerns and potential negative effects on social welfare have been recognized both by regulators~\citep{ohlhausenShouldWeFear2017, bundeskartellamtAlgorithmsCompetition2019, directorate-generalforcompetitioneuropeancommissionCompetitionPolicyDigital2019} and scholars~\citep{harringtonDevelopingCompetitionLaw2018, benekeRemediesAlgorithmicTacit2021, breroLearningMitigateAI2022}.
To develop comprehensive legislation on algorithmic pricing, a thorough understanding of the factors that influence the emergence of collusive strategies is required under assumptions that align with real markets~\citep{calvanoProtectingConsumersCollusive2020}.

Previous research has already shown that \emph{reinforcement learning (RL)} algorithms can engage in tacit collusion in pricing games with infinite time-horizon~\citep{asker2022artificial, calvanoArtificialIntelligenceAlgorithmic2020a, musolff2022algorithmic, klein2021autonomous}.
However, most markets follow some form of periodicity, e.g., seasonality in retail or fiscal years for public companies, which breaks the continuity of the interactions between sellers.
In the markets of perishable goods, hotels, or tickets, the markets only persist until the given sell-by dates and sellers are aware of the finite nature of competition.
Importantly, in the previously investigated infinite time-horizon settings the collusive equilibrium is maintained via punishment strategies, e.g., grim-trigger, but it is not an equilibrium in the finite-horizon case.
This is because these strategies are only credible if sufficient time remains for the punishment to offset short-term gains from deviating from collusion. In the finite-horizon setting, such punishments become unmaintainable as the sell-by date approaches.
However, RL algorithms show the potential to learn collusion through their memory over several episodes interacting against the same opponents.
Additionally, in finite time-horizon markets supplies are often predetermined and limited, therefore, pricing strategies have to consider additional constraints and anticipate future demand to avoid expiring inventory while maximizing total profit.
Both aspects are crucial in many real-world markets. For example, airlines selling tickets between two cities on a certain day have to fill their planes' capacity before departure. However, selling tickets too quickly could lead to a missed opportunity to sell tickets closer to departure time to less price-sensitive consumers, while selling tickets too slowly could result in empty seats.
The added complexity of finite time horizon and inventory constraints results in more complex strategies and interactions between pricing algorithms; therefore, previous results do not immediately hold and further investigation is necessary to develop comprehensive collusion mitigation approaches.

In this work, we aim to contribute to these efforts by extending the analysis of tacit collusion between pricing algorithms to \emph{episodic markets with inventory constraints}. 

In particular, in \Cref{sec:rw}, we give an overview of related literature.
In \Cref{sec:problemstatement}, we define the episodic, finite-horizon pricing problem with inventory constraints as a Markov game, inspired by Airline Revenue Management (ARM), and formalize both competitive (Nash) and collusive (monopolistic) equilibrium strategies. Building on these, we define a measure that quantifies collusion in an observed episode. Notably, our definitions are on the space of pricing strategies instead of price levels at a certain point in time which is the standard in the infinite-horizon setting. This is a significant change in the analysis and a challenge in episodic markets compared to the infinite-horizon case.
In \Cref{sec:collusion_comparison}, we discuss how our model's finite time horizon and inventory constraints change the dynamics of collusion compared to previous work. Reward-punishment schemes cannot extend past the end of the episode, making collusion theoretically impossible (with a backward induction argument), but practically achievable (with imperfect learning agents and long enough episodes).
In \Cref{sec:experiments}, we demonstrate efficient computation of the competitive Nash Equilibrium, a challenging task on its own.
We show that two common deep RL algorithms, Proximal Policy Optimization (PPO) \citep{schulmanProximalPolicyOptimization2017} and Deep Q-Networks (DQN) \citep{mnihHumanlevelControlDeep2015}, learn to collude in our model in two distinct ways that align with the intuition provided in \Cref{sec:collusion_comparison}. We analyze the learned strategies, finding that agents collude while being aware of the competitive best response, and maintain collusion with a reward-punishment scheme. We show that collusion is robust to changes in agent hyperparameters, unless learning targets are made intentionally unstable, in which case agents converge to a competitive best response strategy.
In \Cref{sec:conclusion}, we conclude and discuss future research directions.
\newpage

\section{Related work}\label{sec:rw}
Our work is related to a line of research into competitive and collusive dynamics that emerge between reinforcement learning algorithmic pricing agents in economic games. We refer to \citet{abadaAlgorithmicCollusionWhere2024} for an excellent survey on this topic, and to \Cref{appendix:litreview} for a more detailed literature review.

Recent research most relevant to us focuses on the Bertrand oligopoly, where agents compete by setting prices and using Q-learning.
The main line of research uses Bertrand competition with an infinite time horizon~\citep{calvanoArtificialIntelligenceAlgorithmic2020a}, with follow-up work using DQN~\citep{hettichAlgorithmicCollusionInsights2021}, varying the demand model~\citep{asker2022artificial}, modeling sequential rather than simultaneous agent decisions~\citep{klein2021autonomous}, or an episodic setting with contexts~\citep{eschenbaumRobustAlgorithmicCollusion2022}. 
Findings reveal frequent, though not universal, collusion emergence, often explained by environmental \emph{non-stationarity} preventing theoretical convergence guarantees. Agents consistently learn to charge supra-competitive prices, punishing deviating agents through 'price wars' before reverting to collusion. The robustness of collusion emergence to factors like agent number, market power asymmetry, and demand model changes underscores the potential risks posed by AI in pricing.

Which factors support and impede the emergence of learned collusion remain debated. Some~\citep{waltmanQlearningAgentsCournot2008, abadaArtificialIntelligenceCan2023} argue collusion results from agents `locking in' on supra-competitive prices early on due to insufficiently exploring the strategy space, suggesting a dependence on the choice of hyperparameters. Most studies identifying collusion used Q-learning, with others showing competitive behavior, raising questions about algorithm specificity~\citep{sanchez-cartasArtificialIntelligenceAlgorithmic2022}. However, recent work~\citep{koiralaAlgorithmicCollusionTwosided2024, dengAlgorithmicCollusionDynamic2024} using PPO in ridesharing markets and infinite Bertrand competition respectively, suggests otherwise. We expand on these findings in a more realistic episodic, finite horizon market with inventory constraints using deep %
RL algorithms (PPO and DQN), to manage our model's larger state spaces and dynamic environments.

\section{Problem statement}\label{sec:problemstatement}
We introduce a multi-agent market model for inventory-constrained goods with a sell-by date, such as perishable items, hotel rooms, or tickets, using airline revenue management (ARM) as an example. We show how to model such markets as a Markov game and define a collusion metric based on the profits achieved under perfect competition and collusion.

\subsection{Episodic Markov games}
An \emph{episodic Markov game}~\citep{littmanMarkovGamesFramework1994} is defined by the tuple $(\statespace, \actionspace, \transitionfunc, \rewardfunc, \timehorizon)$ where $\statespace$ represents the common state space shared by all agents, $\actionspace = \actionspace_1 \times \dots \times \actionspace_\numagents$ denotes the joint action space for $\numagents$ agents, $\transitionfunc: \statespace \times \actionspace \to \mathcal{P}(\statespace)$ is the stochastic state transition function, $\rewardfunc_i: \statespace \times \actionspace \to \mathbb{R}$ defines the reward received by agent $i$, and $\timehorizon$ specifies the episode length in discrete timesteps.

At each time step $t$, agents observe the current state $\agentstate_t \in \statespace$ and simultaneously choose actions following their respective time-dependent policies $\pi_{i, t}: \statespace \to \mathcal{P}(\actionspace_i)$. We use $\pi_i$ to denote agent $i$'s vector of policies over time. Each agent's goal is to maximize its cumulative reward over the episode given the game's dynamics,
\begin{subequations}
    \begin{equation*}
        \max_{\pi_i} \sum_{t=1}^\timehorizon \rewardfunc_i (\agentstate_t, \action_t)
    \end{equation*}
    \begin{equation*}
        \textbf{s.t.} \quad \agentstate_{t+1} \sim \transitionfunc(\agentstate_t, \action_t);\quad\action_{j,t} \sim \pi_{j,t}(\agentstate_t).
    \end{equation*}
\end{subequations}

\newpage

The main challenge in finding optimal policies in a Markov game is that agent $i$'s optimization problem depends on the actions chosen by all other agents. In a learning context, where agents optimize their policies simultaneously, this optimization becomes non-stationary and convergence is not guaranteed. For a detailed discussion on the challenges of multi-agent reinforcement learning we refer the reader to a number of surveys~\cite{bucsoniu2010multi, yangOverviewMultiAgentReinforcement2021, gronauer2022multi, wong2023deep}.

\subsection{Markets as episodic Markov games}\label{sec:markovgame_definition}
\looseness=-1
We extend the \emph{Bertrand competition}~\citep{bertrandReviewOf1883} model, where agents compete to sell a common good. In its simplified one-shot setting, sellers choose prices, and consumers react, deciding which quantity to buy from each seller based on some demand function of those prices. In contrast, we model markets where goods can be sold in multiple timesteps $t = 1, \ldots, \timehorizon$ over a finite episode. The Markov game's action space $\actionspace$ consists of the prices agents can set, and an agent's policy $\policy_i$ represents their pricing strategy. Each timestep $t$, agents observe the state $\agentstate_t$ and simultaneously use their policy $\policy_i$ to choose an action in the form of a \emph{price} $\price_{i,t} = \policy_i(\agentstate_t)$, forming the price vector $\price_t = (\price_{1,t}, \ldots, \price_{\numagents,t})$. In the following, we use $p_{i,t}$ for actions instead of $a_{i,t}$ to emphasize that the actions represent prices. 

\looseness=-1
Additionally, we assume that each agent has a finite \emph{capacity} $\capacity_i \in \NN$ of goods that they can sell throughout the episode. At each time $t$, each agent has a remaining \emph{inventory} of tickets $\inventory_{i,t} \in \{0,\ldots,\capacity_i\}$, resulting in an inventory vector $\inventory_t=(\inventory_{1,t}, \ldots, \inventory_{\numagents,t})$.
We define the state of the game at time $t$ as the most recent price vector and current inventory, i.e., $\agentstate_t = (\price_{t-1}, \inventory_t)$.
We motivate this definition of the state by the fact that in the non-episodic setting, most recent prices provide agents sufficient information to learn various strategies including perfect competition and collusion~\citep{calvanoArtificialIntelligenceAlgorithmic2020a, eschenbaumRobustAlgorithmicCollusion2022}.
However, investigating the effect of longer recall is an interesting direction for future research.

With prices chosen, a state transition from time $t$ to $t+1$ occurs: For each agent $i$, the market determines a \emph{demand} $\demand_{i,t}$, the agent sells a corresponding \emph{quantity} $\quantity_{i,t} = \min(\demand_{i,t}, \inventory_{i,t})$ bounded by their inventory, and their inventory is updated to $\inventory_{i,t+1}=\inventory_{i,t}-\quantity_{i,t}$. With our choice of demand function (cf. \cref{sec:mnl-definition}), this transition to the next period's state $\agentstate_{t+1} = (\price_t, \inventory_{t+1})$ is deterministic. Finally, each agent receives their profit as a \emph{reward} $\reward_{i,t}:=\reward_{i}(\agentstate_t, \price_t) = (\price_{i,t}-\cost_i)\quantity_{i,t}$, with $\cost_i$ their constant \emph{marginal cost} per good sold.

\subsection{Application to airline revenue management}\label{sec:airline_example}
To motivate the episodic Markov game framework, we consider the Airline Revenue Management (ARM) problem. In ARM, agents represent airlines competing to sell a fixed number of seats on a direct flight (also called a \emph{single-leg} flight) between two cities on the same day. The problem is naturally episodic; episodes start when the flight schedule is announced and end at departure, i.e., the sell-by date of the tickets. Furthermore, each airline is constrained by the capacity of their respective aircraft. We consider each route on each day to form a single independent market. Expanding our model to connecting (\emph{multi-leg}) flights, several flights on the same day, cancellations, and overbooking promises interesting future work.
This market is a great example with fierce competition, a history of tacit collusion~\citep{borenstein1994competition}, real-time public information on offered ticket prices and inventories via Global Distribution Systems (GDS), and early adoption of dynamic pricing algorithms~\citep{koenigsbergEasyJetAirlines2004}\footnote{Adoption of dynamic pricing algorithms in this industry has historically been limited to low-cost carriers, due to established carriers heavily depending on legacy systems and data-driven forecasting models. See lit. review in \Cref{appendix:litreview}.}.

\subsection{Demand model}\label{sec:mnl-definition}

\looseness=-1
We employ a modified \emph{multinomial logit (MNL)} demand model, commonly used in Bertrand price competition~\citep{calvanoArtificialIntelligenceAlgorithmic2020a, eschenbaumRobustAlgorithmicCollusion2022, dengAlgorithmicCollusionDynamic2024}, to simulate the probability of a customer choosing each agent's product, ensuring demand distribution among all agents rather than clustering on the best offering.
The normalized \emph{demand} for agent $i$'s good in period $t$ is
\begin{align*}
    \demand_{i,t} = \frac{\exp\bigl((\quality_i-\price_{i,t})/\horizontaldiff\bigr)}{\sum_{j \in \agentset^a_t} \exp\bigl((\quality_j-\price_{j,t})/\horizontaldiff\bigr) +\exp(\quality_0/\horizontaldiff)} \in (0,1),
\end{align*}
where $\agentset^a_t:= \{j \in \agentset\ |\ \inventory_{j,t} > 0\}$, $\quality_i$ is agent $i$'s good's quality, $\quality_0$ is the quality of an outside good for vertical differentiation, and $\horizontaldiff$ is the horizontal differentiation scaling parameter.
The quantity demanded from agent $i$ at time $t$ is then defined as $q_{i,t} = \min \{ \lfloor \demandscalingfactor d_{i,t} \rfloor, x_{i,t} \}$, scaling demand with a factor $\demandscalingfactor \in \NN$ and rounding to the nearest integer to account for the sale of goods in whole numbers.
We incorporate \emph{choice substitution}, or \emph{demand adaptation}, by summing only over agents with available inventory $\agentset^a_t$. If an agent is sold out, demand shifts to those with remaining inventory, preventing the sold-out agent's actions from affecting the demand and rewards of others. 

\subsection{Measuring collusion and competition}
We measure the collusion of an observed episode and agent strategies on a scale from $0$ (\emph{competitive}) to $1$ (\emph{collusive}).
First, we establish the two extremes in the Markov game as the competitive Nash equilibrium and the monopolistic optimum that we can later use as reference points for collusion.

\begin{definition}[Competitive \& collusive solutions]
\label{def:comp_coll_solutions}
    A collection of agent policies $(\pi_1, \dots, \pi_\numagents)$ is called 
    \begin{itemize}
        \item \emph{Competitive}, or \emph{Nash equilibrium}, if no agent $i$ can improve their expected episode profit $\EE_\policy[\Sigma_{t=1}^\timehorizon \reward_{i,t}]$ by unilaterally picking a different policy given fixed opponent policies. %
        \item \emph{Collusive}, or \emph{monopolistic optimum}, if it maximizes expected collective profits, $\EE_\policy[\Sigma_{i=1}^\numagents \Sigma_{t=1}^\timehorizon \reward_{i,t}]$. %
    \end{itemize}
\end{definition}

As we argue theoretically in \cref{sec:collusion_comparison} and show experimentally in \cref{sec:gnep-def}, both admit solutions that feature constant prices across an episode, which we call $\price^N$ and $\price^M$ for the Nash and monopoly cases, respectively. In our model, the collusive prices $\price^M$ are higher than the competitive prices $\price^N$, and the same holds for the correspondingly achieved profits $\reward^M$ and $\reward^N$. At the Nash equilibrium, both unilaterally increasing or decreasing one's price reduces profits. However, if all agents jointly increase prices, the increase in margin outpaces the decrease in (MNL) demand, leading to increased profits for everyone.
Building on these two solutions, we define a measure for collusion.

\begin{definition}[Collusion measure]
    \label{def:collusion_measure} 
    We define agent $i$'s \emph{episodic profit gain} as
    $$\profitgain_{i,e} := \frac{1}{\timehorizon} \sum_{t=1}^{\timehorizon} \frac{\bar{\reward}_{i,t}-\reward_{i,t}^N}{\reward_{i,t}^M-\reward_{i,t}^N}.$$

    The \emph{episodic collusion index} is measured as the generalized mean of the individual episodic profit gains, i.e.,
    $$\collindex_e := \Biggl(\frac{1}{\numagents} \sum_{i=1}^\numagents \profitgain_{i,e}^\gamma \Biggr)^{\frac{1}{\gamma}} $$
    indicating a competitive or collusive outcome at 0 or 1, respectively.
\end{definition}

\looseness=-1
The generalized mean interpolates the arithmetic mean (i.e., average) and geometric mean, which are obtained by setting $\gamma=1$ and $\gamma=0$ respectively. We use $\gamma=0.5$ for our collusion index. Our reason is that the geometric mean has an advantage against the simple average used in previous studies \citep{calvanoArtificialIntelligenceAlgorithmic2020a, eschenbaumRobustAlgorithmicCollusion2022}, as it more strongly penalizes unilateral competitive defections in a collusive arrangement. However, it interprets any outcome where at least one agent achieves only competitive, or even sub-competitive profits (defining the measure via clamping negative profit gains to zero) as fully competitive, even if others prices above the competitive level and achieve considerable supra-competitive profits. The generalized mean provides a good middle ground. To better interpret negative values, we replace $\profitgain_{i,e}^\gamma$ with $\sgn(\profitgain_{i,e}) |\profitgain_{i,e}|^\gamma$. See \Cref{appdx:means_comparison} for a comparison of means. Ultimately, how to aggregate the individual profit gains is a subjective question with trade-offs that depend on which outcomes one wants to differentiate the best. E.g., the following outcomes $(\profitgain_{1,e}, \profitgain_{2,e})$ of $(0.1, 0.1)$, $(0, 0.2)$ or $(-0.1, 0.3)$ have the same average episodic profit gain, but quite different agent behavior and implications on consumer welfare, especially if agents' qualities, costs, and thus equilibrium profits, are not symmetric. Exploring alternative measures, which could be inspired by social choice theory, is a promising avenue for future research.

\section{The collusive strategy landscape}
\label{sec:collusion_comparison}
In this section, we discuss how our model's episodic nature and finite inventory significantly affect the strategies for establishing and maintaining learned tacit collusion compared to the previously considered infinite horizon setting.
It is common economic intuition (e.g.,~\citep{harringtonDevelopingCompetitionLaw2018}) that in order to maintain collusive agreements, agents need to remember past actions and have mechanisms to punish those who deviate from the agreed-upon strategy\footnote{Recent work~\citep{arunachaleswaranAlgorithmicCollusionThreats2024} suggests that there can exist stable, collusive equilibria of strategies that do not encode threats. They show that near-monopoly prices can arise if a first-moving agent deploys a no-regret learning algorithm, and the second agent subsequently picks a non-responsive pricing policy.}.
Standard punishment strategies include a temporary or permanent shift to a competitive price level after the deviation is detected which results in lower profits for all firms.
It has been well documented that learning algorithms converge to these strategies in the infinite horizon setting~\citep{calvanoArtificialIntelligenceAlgorithmic2020a,hettichAlgorithmicCollusionInsights2021,dengAlgorithmicCollusionDynamic2024}.
Such strategies are only credible as long as sufficient time and supply is available for the punishment to offset the short-term gains from a deviation.
These conditions are not always met in our settings that lead to new collusive strategies.

\paragraph{Infinite horizon games} These settings allow for deriving unique competitive and collusive equilibrium price levels through implicit formulas with the most commonly used Bertrand competition models. They provide the most room for collusive strategies to emerge and sustain since there is no time constraint for a punishment strategy's credibility. Typically, stable collusion manifests in two forms. 
First, \emph{reward-punishment schemes:} Agents cooperate by default and punish deviations. A deviating agent is punished by others charging competitive prices, thereby removing the benefits of collusion temporarily, until the supra-competitive prices are reinstated. This dynamic involves agents synchronizing over rounds to restore higher price levels after a deviation.
This pattern can be observed as fixed, supra-competitive prices and verified by forcing one agent to deviate and recording everyone else's responses.
Second, \emph{Edgeworth price cycles:} This pattern involves agents sequentially undercutting each other's prices until one reverts to the collusive price, prompting others to follow, restarting the undercutting cycle~\citep{klein2021autonomous}.

\paragraph{Episodic games} In comparison to the infinite horizon setting, collusive strategies can now emerge in two distinct ways. First, through \emph{intra-episode} action-based communication, where agents gradually raise their prices through signaling within a single episode. Second, through training \emph{across many episodes}, where agents eventually learn policies that implement collusive pricing immediately from the start of each new episode. The latter form is prevalent in oligopolistic settings and possibly explained by learners overfitting their strategies to familiar opponents. When faced with new opponents, collusive agents initially play competitively before reestablishing collusion through continued learning~\citep{eschenbaumRobustAlgorithmicCollusion2022}. 
This robustness result suggests that firms aiming to collude can pre-train their pricing agents separately, needing only (likely legal) alignment on the high-level training setups (e.g., algorithm classes, observation modeling, exploration schedule).
In our experiments in \Cref{sec:exp_strategy_analysis}, we observe evidence of both types of collusion. 

\looseness=-1
The finite time horizon restricts collusive potential by limiting the efficacy of reward-punishment schemes used in infinite-horizon games to maintain collusion. In a one-shot game ($\timehorizon=1$) in our Bertrand setting, there exists a unique Nash equilibrium at the competitive price level, as unilateral deviation from collusive prices is profitable and future punishment is impossible. In the finite horizon case ($\timehorizon>1$), the same logic applies at the final period ($t=\timehorizon$), such that any Nash equilibrium strategy will price competitively in the last timestep. By induction from $t=\timehorizon$ backwards, this argument extends to all periods $t=\timehorizon-1, \ldots, 1$, defining a unique Nash equilibrium where agents compete throughout the episode. Does this mean that collusion in episodic games is impossible? No: If agents remember past interactions across episodes, deviations can be punished in future episodes. 
Surprisingly, our experiments in \Cref{sec:experiments} show that even without cross-episode memory, learning agents in sufficiently long episodes can converge to collusive strategies of the signaling, stable or cyclic kind. 
We observe that some agents learn to play collusively at episode start and defect toward the end, suggesting that discovering the full backward induction argument through (often random) exploration is unlikely enough in practice.

\paragraph{Episodic, inventory-constrained model}
Inventory constraints significantly complicate the state and strategy space by making the reward achieved from a pricing strategy dependent on inventory levels. Determining the competitive and collusive price levels becomes more complex because the solution formulas from the Bertrand or Cournot settings require smoothness or convexity assumptions that no longer hold, preventing the standard uniqueness proofs. We approach finding a Nash equilibrium by modeling each episode as a simultaneous-move game where agents set entire price vectors before the episode starts for the complete episode. We provide further details in \Cref{sec:gnep-def}. We solve the resulting generalized Nash equilibrium problem numerically and prove that its solutions are Nash equilibria in our Markov game. We find that in our model, both the competitive and collusive solutions consist of repeating their prices from the one-period equivalents $\timehorizon$ times.
If agents discount future rewards, both equilibria shift to lower prices and higher profits early in the episode and vice versa toward its end. In addition, price levels remain distinct even with strict inventory constraints. Due to the difficulty in predicting or interpreting observed behavior in this complex setting, we see value in analyzing different types of learners as part of future work.

\section{Experiments}\label{sec:experiments}
In \cref{sec:gnep-def} we first show how to find the competitive and monopolistic price levels needed to calculate the collusion measure defined in \Cref{def:collusion_measure}, and how they change under different inventory constraints. Then, we show that PPO~\citep{schulmanProximalPolicyOptimization2017} and DQN~\citep{mnihHumanlevelControlDeep2015}, two commonly used deep RL algorithms, can learn to collude in our episodic model. Finally, we analyze their learned strategies and their dependence on hyperparameters.

\subsection{Obtaining competitive and collusive equilibrium prices}\label{sec:gnep-def}
Previous works' Bertrand settings use analytic formulae to compute Nash equilibrium and monopolistic optimum price vectors $\price^N$ and $\price^M$ for single-period cases. However, a closed-form solution is not available for our problem setting. We therefore use numerical methods to calculate the competitive and collusive solutions as defined in \Cref{def:comp_coll_solutions} and use these values to define the collusion measure in \Cref{def:collusion_measure}.

First, we calculate the profits and prices in the monopolistic (perfectly collusive) setting by assuming a central optimizer who chooses prices for all agents maximizing the total profit. Second, to calculate the same for the competitive Nash equilibrium, we model an entire episode as a \emph{simultaneous-move game (SMG)}, where all agents $i$ must simultaneously decide all $\timehorizon$ prices in their vector $p_i = (p_{i,1}, \ldots, p_{i,\timehorizon})$ before an episode begins. Let $\price = (\price_1,\ldots,\price_\numagents)$ encompass all agents' price vectors, with $\price_{-i}$ representing all agents' vectors except $i$'s.
The solution to this SMG is then a Generalized Nash Equilibrium defined as follows.

\begin{definition}\label{eq:gnep}
    The \emph{Generalized Nash Equilibrium Problem (GNEP)} consists of finding the price vector ${\price^\ast = (\price_1^{\ast},\ldots, \price_\numagents^{\ast})}$ such that for each agent $i$, given $\price_{-i}^{\ast}$, the vector $\price_{i}^{\ast}$ solves the following inventory-constrained revenue maximization problem
    \begin{align*}
        \begin{split}
            \max_{p^{(i)}} \quad  &\sum_{t=1}^\timehorizon (\price_{i,t} - \cost_i) \lfloor\lambda\demand_{i,t}\rfloor \\
            \text{subject to} \quad &\sum_{t=1}^\timehorizon \lfloor\lambda\demand_{i,t}\rfloor \leq \capacity, \quad p_i \geq 0.
        \end{split}
    \end{align*}
\end{definition}

The solution price vector $\price^\ast$ can be interpreted as the \emph{actions} of a set of agent policies playing an episode of the Markov game. The following lemma shows that a set of policies that result in the price vector $\price^\ast$ form a Nash Equilibrium in the Markov Game.

\begin{lemma}
    Given a Markov Game with deterministic transitions, let $\price^\ast = (\price_1^{\ast},\ldots,\price_\numagents^{\ast})$ be the solution to \Cref{eq:gnep} and define $\policy^\ast = (\policy_1^\ast, \ldots, \policy_\numagents^\ast)$, as $\policy_i^\ast(\agentstate_t) = \price^{\ast}_{i,t}$ for all $i$, $t$, and $\agentstate_t \in \statespace$. Then $\policy^\ast$ is a Nash equilibrium in the Markov Game. 
\end{lemma}

The full proof can be found in \Cref{appendix:proof-lemma1}. Details of our numerical approach to solving the GNEP are found in \Cref{appendix:gnep-solving}.

\begin{figure}
    \centering
    \includegraphics[width=0.8\linewidth]{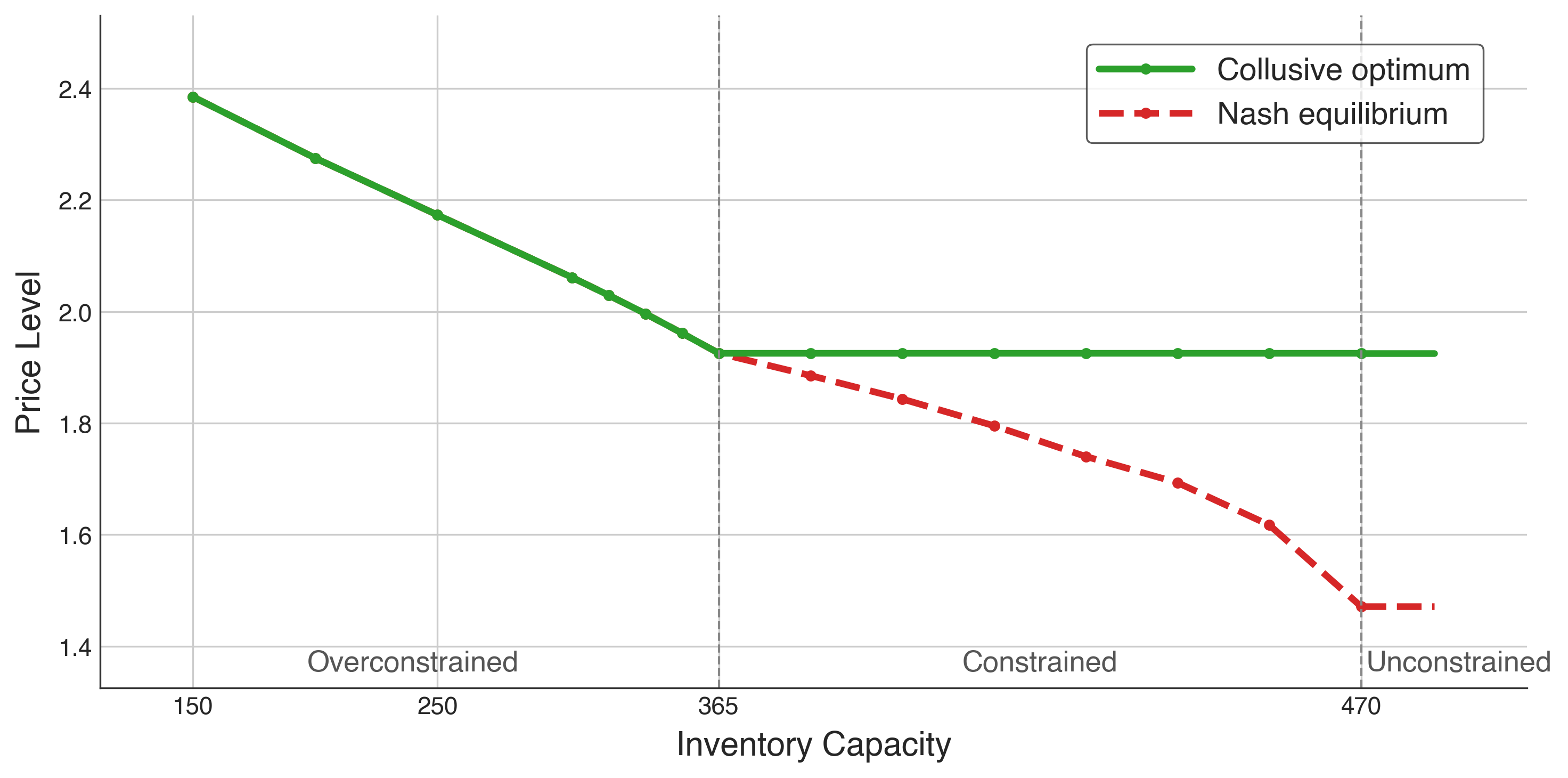}
    \caption{One-period equilibrium price levels as a function of inventory capacity for two equally constrained agents.}
    \label{fig:equilibria}
\end{figure}

\looseness=-1
Without discounting, the episodic equilibrium price vectors repeat the single-period equilibrium with the same parameters $\timehorizon$ times. \Cref{fig:equilibria} shows how inventory constraints affect market dynamics. When inventories exceed the demand at the competitive equilibrium, the equilibria correspond to the unconstrained setting. As inventories shrink, the competitive price level rises, as it is harder for firms to undercut and profit from the increased demand. When inventory size matches the demand at the collusive price, the collusive and competitive price levels converge. Further tightening of constraints pushes both coinciding prices higher.
In our experiments we choose the constraint's value between the two extremes to allow for differentiation between competitive and collusive behavior and a well-defined collusion index, and investigate the effect of the inventory size on learned collusion in \Cref{sec:hyperparams}.

\subsection{Model parameters}
\label{sec:model_parameters}
We evaluate the potential for RL algorithms to collude in our model using a duopoly situation with two agents\footnote{Two agents suffice to demonstrate learned collusion in the finite horizon game and the impact of inventory constraints. A duopoly is a reasonable assumption in the ARM domain, as many routes are dominated by 2-3 airlines. For $\numagents>2$ agents, \cite{abadaArtificialIntelligenceCan2023, hettichAlgorithmicCollusionInsights2021} show collusion indeed diminishes due to exponential growth in joint policy space hindering joint exploration, but does not fully disappear.}%
. We use either of two popular algorithms, namely Deep Q-Networks (DQN)~\citep{mnihHumanlevelControlDeep2015} and Proximal Policy Optimization (PPO)~\citep{schulmanProximalPolicyOptimization2017} for learning without weight-sharing between agents.
Agents represent identical firms, sharing the same qualities $\quality_i = 2$, marginal costs $\cost = \cost_i = 1\ \forall i$, a horizontal differentiation factor of $\horizontaldiff=0.25$, an outside good quality of $\quality_0 = 0$, and a demand scaling factor of $\demandscalingfactor=1000$.
For the main results presented in \Cref{sec:experiment_learning_process} and \Cref{sec:exp_strategy_analysis}, we set the inventory constraints to $440 \cdot \timehorizon$ and the episode length $\timehorizon=20$.

Due to the symmetry between agents, Nash and monopolistic price levels are identical for both of them, and the price levels and the corresponding demands are $\price^N=1.693, \price^M=1.925$ and $\demand^N=440 \demand^M=365$ for our inventory constrained case.
Agents choose prices from a discretized interval $[\price^N - \xi(\price^M-\price^N), \price^M + \xi(\price^M-\price^N)]$ with 15 steps and $\xi=0.2$, such that the competitive and collusive actions correspond to $\action^N=2$ and $\action^M=12$ respectively.
In particular, the price range for our setting is $[1.693, 1.925]$.
In \Cref{sec:appendix_price_action}, we provide further results on experiments with a price range defined with the unconstrained Nash equilibrium prices to demonstrate that agents are still capable of learning collusion and their actions quickly converge to the price range defined with the constrained Nash equilibrium prices.

\subsection{Training setup}
\looseness=-1 %
We train our algorithms by playing $1000$ and $50,000$ episodes for PPO and DQN, respectively, and updating weights after every episode for PPO or every fourth for DQN. We train $100$ pairs of PPO or DQN on unique random seeds ($40$ for the boxplots). After training, we analyze each agent pair by observing their play in a single episode. This joint training aligns with previous work~\cite{calvanoArtificialIntelligenceAlgorithmic2020a, koiralaAlgorithmicCollusionTwosided2024} and real market situations, where firms learn while competing, updating pricing strategies based on market success.  
Solid lines and shaded areas in our plots represent the averages and standard deviations of their metrics.
For our DQN agent, we use %
\emph{epsilon-greedy} exploration with an exponentially decaying epsilon, while the PPO agent anneals its entropy coefficient to similarly reduce %
exploration over time. For evaluation episodes, DQN uses a fully greedy action selection. We normalize the 
rewards during training to the interval $[0,1]$ based on %
minimum and maximum possible values. This makes training slightly more stable. However, collusion is still achieved with unnormalized rewards.
A full description of the hyperparameters used for DQN and PPO can be found in \Cref{appendix:dqn_ppo_params}.
We use the JAX framework on a custom codebase built on~\citep{pax}.
Our experiments were run on a compute cluster on a mix of nodes with each run using at most four vCPU cores, 8GB of RAM, and either a NVIDIA T4 or NVIDIA V100 GPU. However, a single run can be done on a consumer laptop (Apple M1 Max, 32GB RAM) in under one hour.

\subsection{Analysis of learning process}
\label{sec:experiment_learning_process}

\begin{figure}[h]
    \centering
    \includegraphics[width=\linewidth]{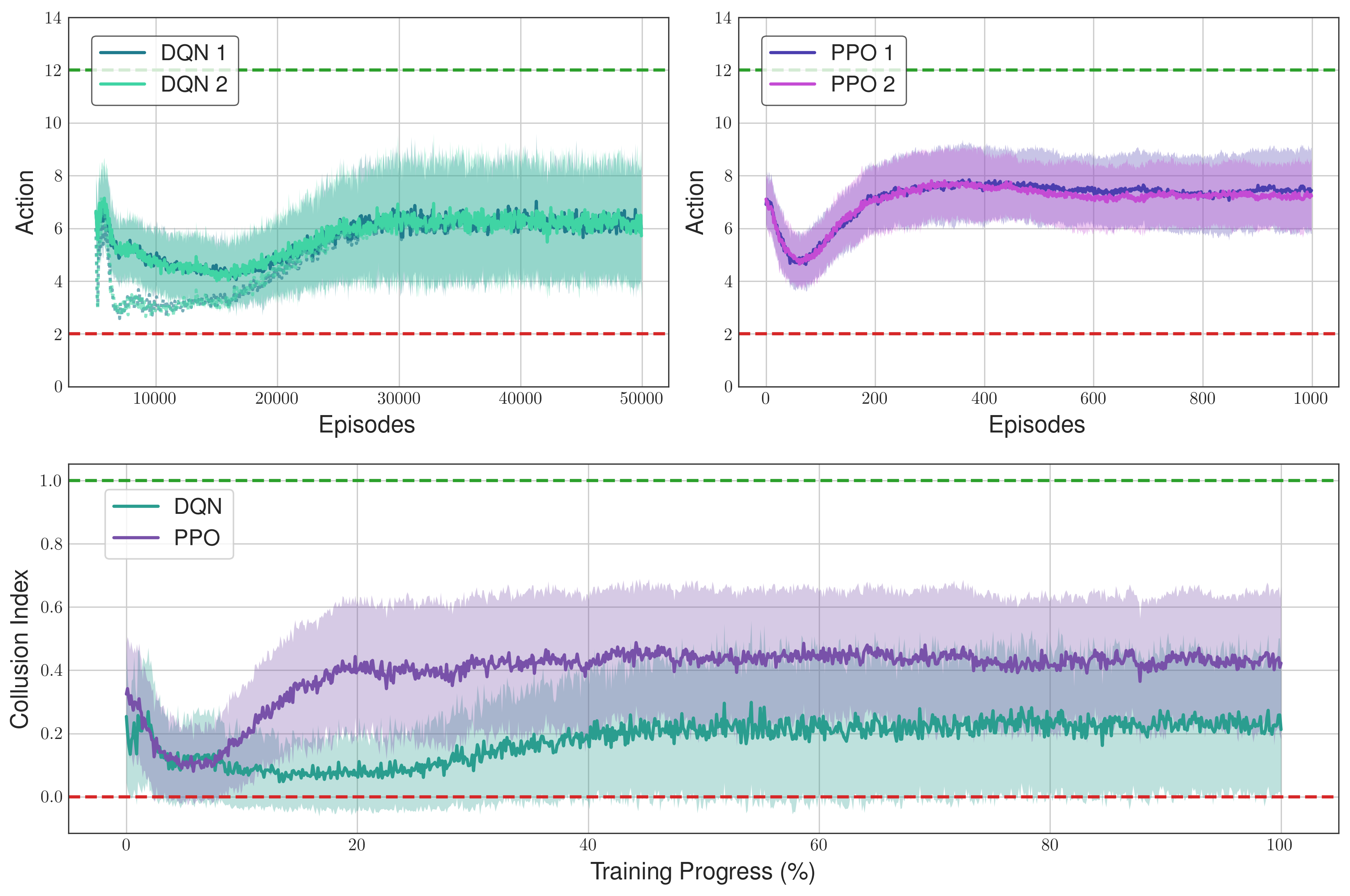} %
    \caption{Evolution of training two DQN and two PPO agents in our model, showing average agent actions per episode (DQNs top left, PPOs top right) and collusion index (bottom) with collusive and competitive actions indicated with the green upper and red lower dashed lines respectively. In DQN's training plot, the dotted lines are the greedy actions that DQN would have chosen. Both DQN and PPO first converge to competition before gradually rising toward collusion.}
    \label{fig:DQN_and_PPO_collusion}
\end{figure}

\Cref{fig:DQN_and_PPO_collusion} shows two training runs for DQN and PPO agents. For both algorithms, agents quickly converge to each other and to competition as their learning targets are initially unstable, with high epsilon (DQN) and entropy (PPO) forcing random actions. This makes it hard for agents to adapt to their opponent's underlying policy and leads to them learning the best-response strategy against a random opponent, playing competitively. As the exponentially decaying epsilon and entropy curves flatten and the agents face an increasingly predictable opponent that they can adapt to, they begin colluding. Prices rise gradually and jointly before leveling off at a collusive level. PPO converges in both much fewer episodes and achieves higher levels of collusion, with an average collusion index of $\collindex_e=0.43$ over the last $10\%$ of episodes, compared to DQN's $\collindex_e=0.23$. 
These values, lower than in prior studies in the standard Bertrand setting~\citep{calvanoArtificialIntelligenceAlgorithmic2020a, hettichAlgorithmicCollusionInsights2021, dengAlgorithmicCollusionDynamic2024}, highlight the greater challenge of collusion in our more complex model.
Regulatory efforts could focus on the gradual increase in prices to mitigate algorithmic collusion, which we consider to be an interesting direction for future work.

\newpage

\subsection{Analysis of collusive strategies}
\label{sec:exp_strategy_analysis}
\begin{figure}[hb]
    \centering
    \includegraphics[width=\linewidth]{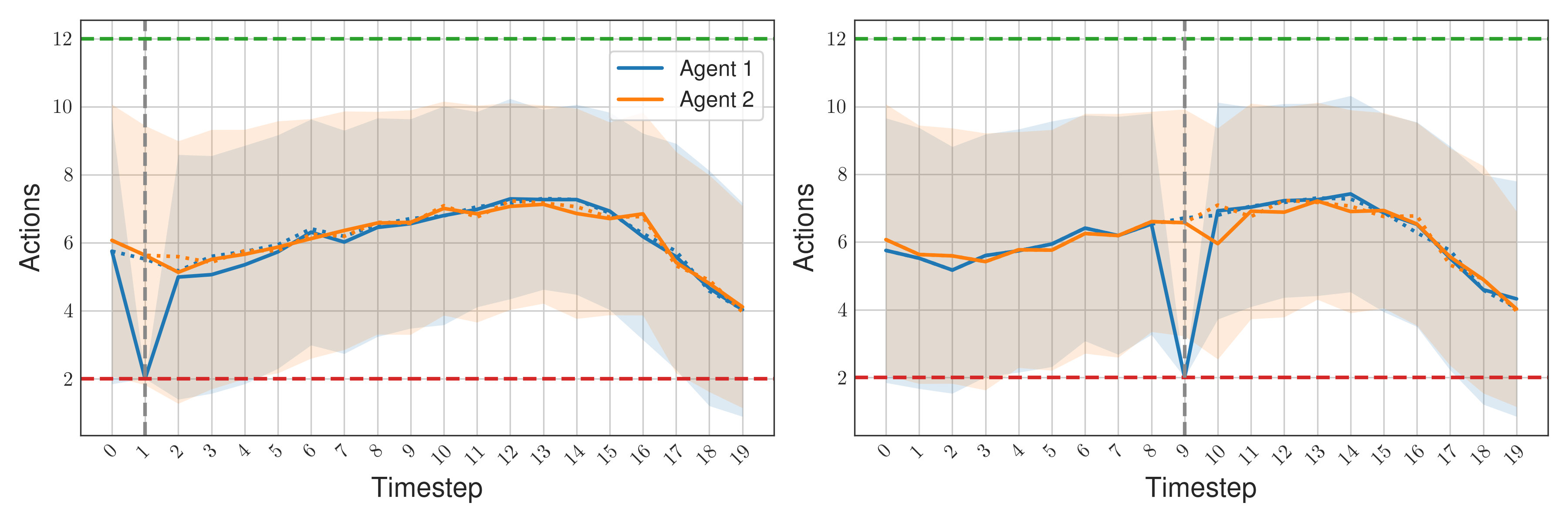}
    \caption{Behavior of two DQN agents during an episode after forcing one agent to deviate at time $t=1$ and $t=9$ respectively. Dotted lines indicate evolution without deviation. Deviations provoke a competitive reaction, with both agents quickly returning to collusion.}
    \label{fig:forced_deviation_DQN}
\end{figure}

After training, we simulate the agents in an evaluation episode (\Cref{fig:forced_deviation_DQN}). We focus on DQN here, discussing PPO in \Cref{sec:PPO_analysis}. Our DQN agents show behavior that slowly rises in collusiveness until both agents defect near the end of the episode. This suggests that the agents are capable of learning that late defection cannot be punished, while not fully applying the backward induction argument from \Cref{sec:collusion_comparison}. The rise in collusion at the beginning of the episode suggests a capability of establishing \emph{intra-episode} collusion, with the gradual, mutual price increase acting as a form of signaling.
In \Cref{sec:appendix_unconstrained_DQN}, we show results without inventory constraints, where the agents' price curve is flatter, suggesting a strategy more based on solidified collusion over multiple episodes.

To analyze the nature of the learned strategy, we force one agent to deviate at a certain timestep and record the response by both agents similarly to~\citet{calvanoArtificialIntelligenceAlgorithmic2020a}. Interestingly, deviation produces only a small reaction by the competing agent, while the deviating agent quickly returns to near their collusive level. With a deviation at time $t=1$ or at $t=9$, the impact on overall episode profits is negligible for both agents, with the deviating agent breaking even and the non-deviating agent losing only $0.2\%$ profit overall. We refer the reader to \Cref{sec:appendix_deviation_analysis} for details.

\begin{figure}[htb]
        \centering
    \includegraphics[width=\linewidth]{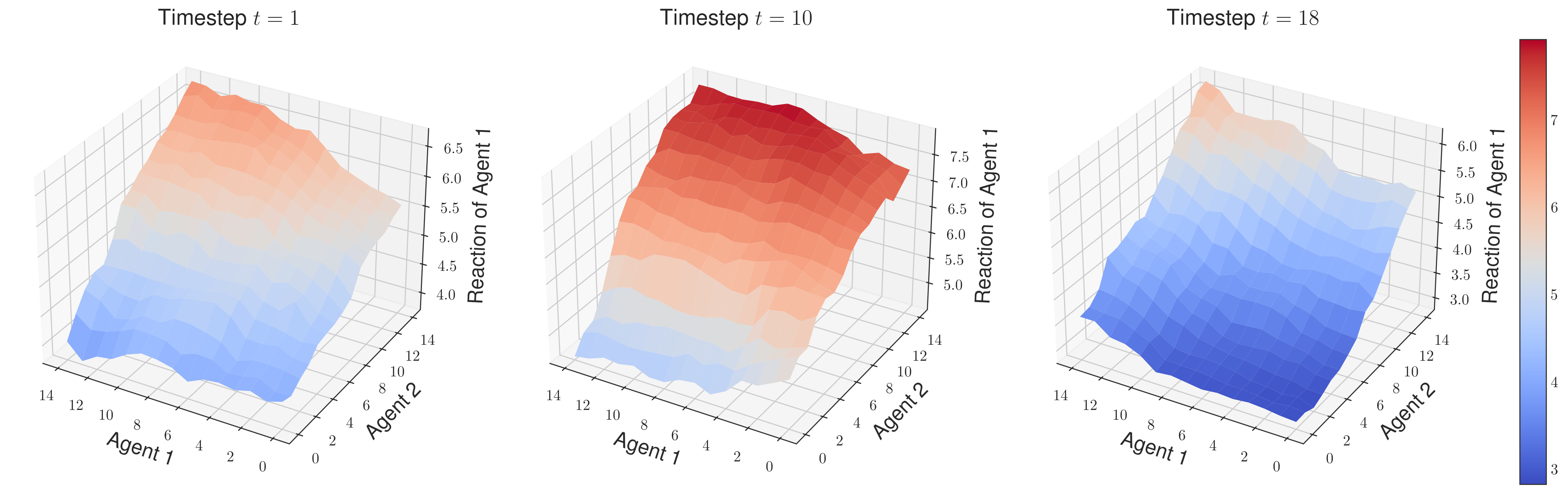}
    \caption{The surfaces show a DQN agent 1's learned best response under their greedy policy (i.e., the action with the highest Q-value) to a state given by both agent's prices (x- and y-axes), timestep and symmetric remaining inventory level.}
    \label{fig:DQN_response_surface}
\end{figure}

\looseness=-1
\Cref{fig:DQN_response_surface} shows the best-response surfaces of the first agent at different points in the episode, with a remaining inventory linearly interpolated from full to none over the episode (corresponding to the agents' evaluated strategy) and averaged over $100$ trained agent pairs. We make two observations. First, the agent always punishes opponent deviations by pricing lower than the previous price level.
Second, at the beginning and end of the episode, the agent's best-response surface shows some symmetry indicative of more competitive behavior. There, the agent will react to their own deviations by pricing even lower in consecutive periods, anticipating a `price war'. During the middle of the episode, the agent instead returns to previous or even higher collusion levels after own defections, signaling cooperation, and punishes opponent deviations with slight undercutting. Near the end of the episode, they shift to more competitive behavior, punishing deviations more strongly.
This topology suggests that if both agents start near the competitive equilibrium, they will both react in a way that jointly `climbs the hill' to collusion, leveling out at an action of roughly $7$ as indicated by the flat top. The second agent behaves similarly.
These results suggest that DQN agents are well aware of competitive strategies and choose to collude in a robust way reliant on rewards and punishments.
\Cref{appendix:observability,,appendix:asymmetric_inventory} contain results for uneven inventory constraints and limiting observability of opponent inventory and time, neither of which significantly hinder the emergence of collusion.

\subsection{Hyper- and environment parameters}\label{sec:hyperparams}
\looseness=-1
We analyze the impact of changing agent hyperparameters and environment characteristics on the convergence and collusive tendencies of DQN and PPO agents. We show comparisons for agent learning rate, inventory constraint, and episode length here, with additional results deferred to \Cref{sec:appendix_more_hparams}. To judge the convergence of two agents toward each other throughout the training run, we use the following metric:
\begin{equation*}
    \frac{1}{0.1\numepisodes}\sum_{e=0.9\numepisodes}^{\numepisodes} \frac{1}{T} \sum_{t=1}^T \frac{|\price_{0,t}-\price_{1,t}|}{\price^M-\price^N}
\end{equation*}
adapted from~\citet{dengAlgorithmicCollusionDynamic2024}, where $\numepisodes$ is the number of training episodes. It takes the average difference of both agents' prices across an episode relative to the width of the Nash-monopolistic price interval. Values below $0.2$ are interpreted as converged.

In our analysis, we vary single parameters from the reference setup described in \Cref{sec:model_parameters}, train agents on $40$ different seeds, and for each parameter value, record the distribution of convergence metric and collusion index over those seeds, averaged over the last $10\%$ of training run episodes.

\begin{figure}[htb]
    \begin{subfigure}[t]{\linewidth}
        \centering
        \includegraphics[width=0.75\linewidth]{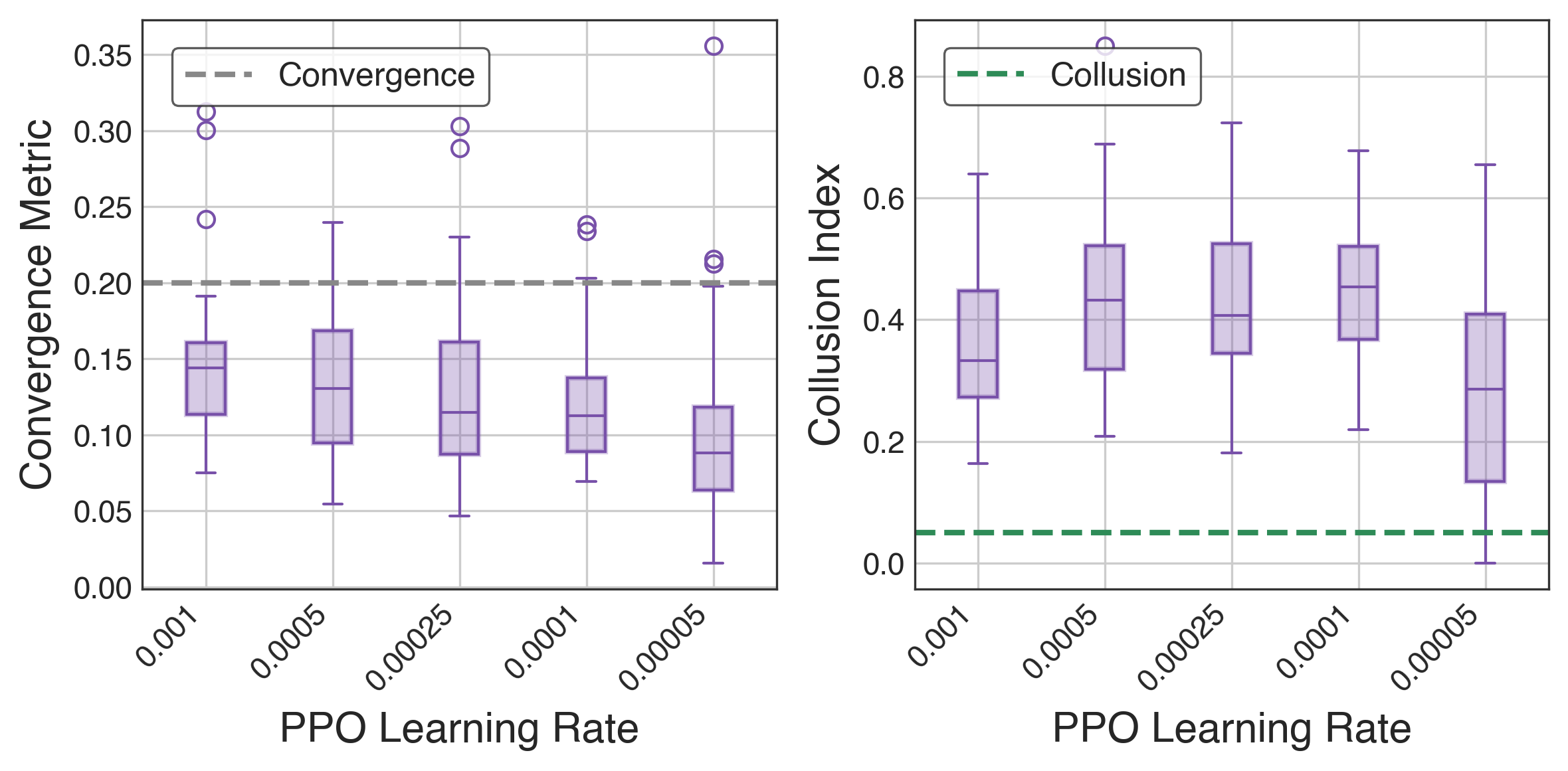}
        \label{fig:DQN_PPO_boxplots:a}
    \end{subfigure}
    \begin{subfigure}[t]{\linewidth}
        \centering
        \includegraphics[width=0.75\linewidth]{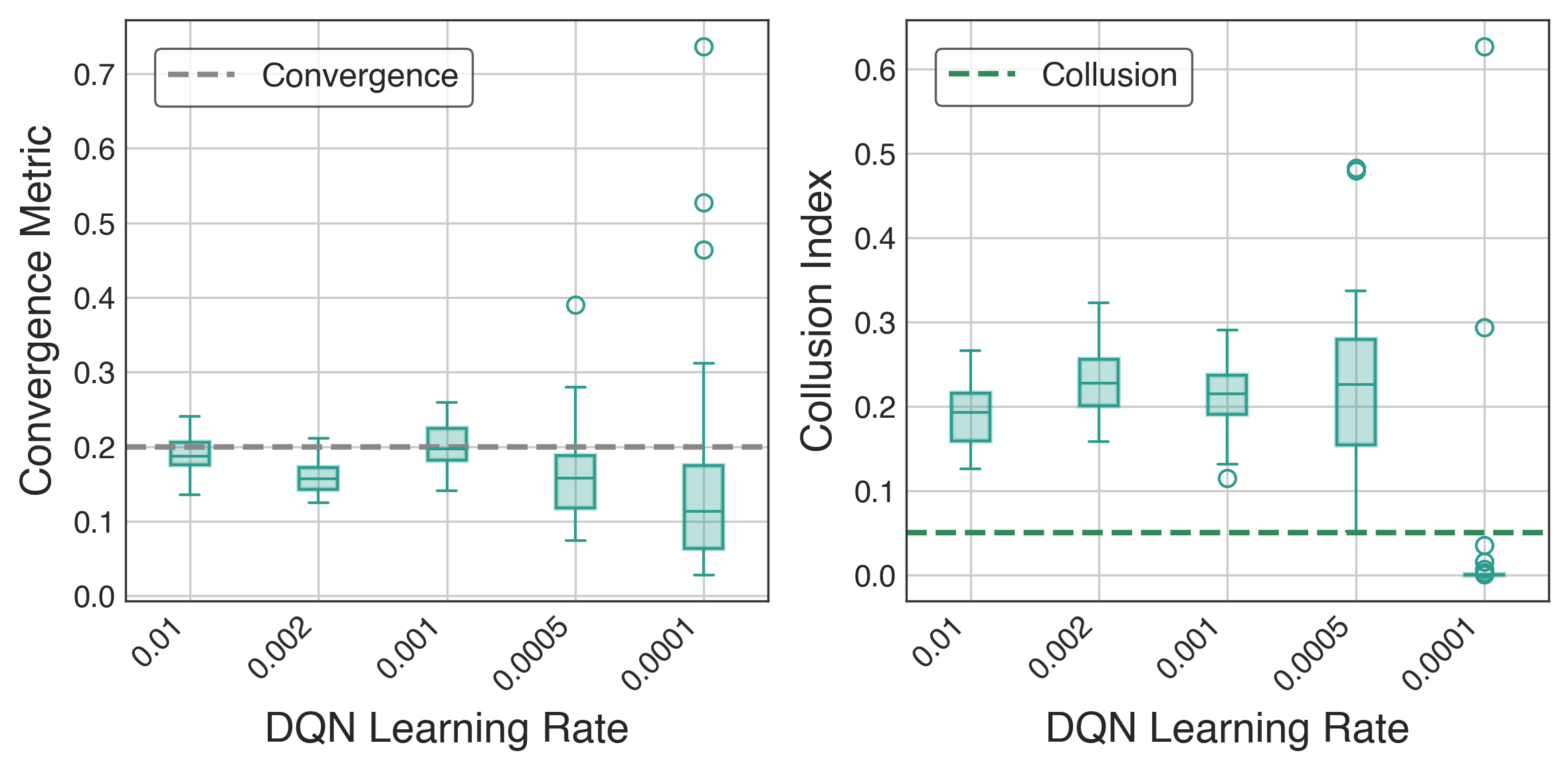}
        \label{fig:DQN_PPO_boxplots:b}
    \end{subfigure}
    \label{fig:DQN_PPO_learningrate_boxplots}
    \caption{Convergence and collusion metrics for DQN and PPO training runs with varied learning rate. Collusion is robust against varying (yet sufficiently large) learning rate.}
\end{figure}

Learning rate is perhaps the most important agent parameter, as it regulates the impact of all other agent parameters. 
\Cref{fig:DQN_PPO_learningrate_boxplots} demonstrates that both PPO and DQN agents achieve better convergence and increased tendency to compete at lower learning rates.
The reduced ability to adapt to an opponent's strategy still allows agents to learn the opponent-independent best-response of competition at initial training episodes, but attempts to establish the gradual, mutual increase in price seen in \Cref{fig:DQN_and_PPO_collusion} happen more rarely and revert to competition more often. A higher learning rate does not translate to more likely collusion, as the increased ability to adapt to an opponent is balanced by the potential to overreact to the opponent's random actions. Overall, collusion and convergence appear to be robust to moderate changes in learning rate.

\begin{figure}[ht]
    \begin{subfigure}[t]{\linewidth}
        \centering
        \includegraphics[width=0.9\linewidth,trim={0 5pt 0 0},clip]{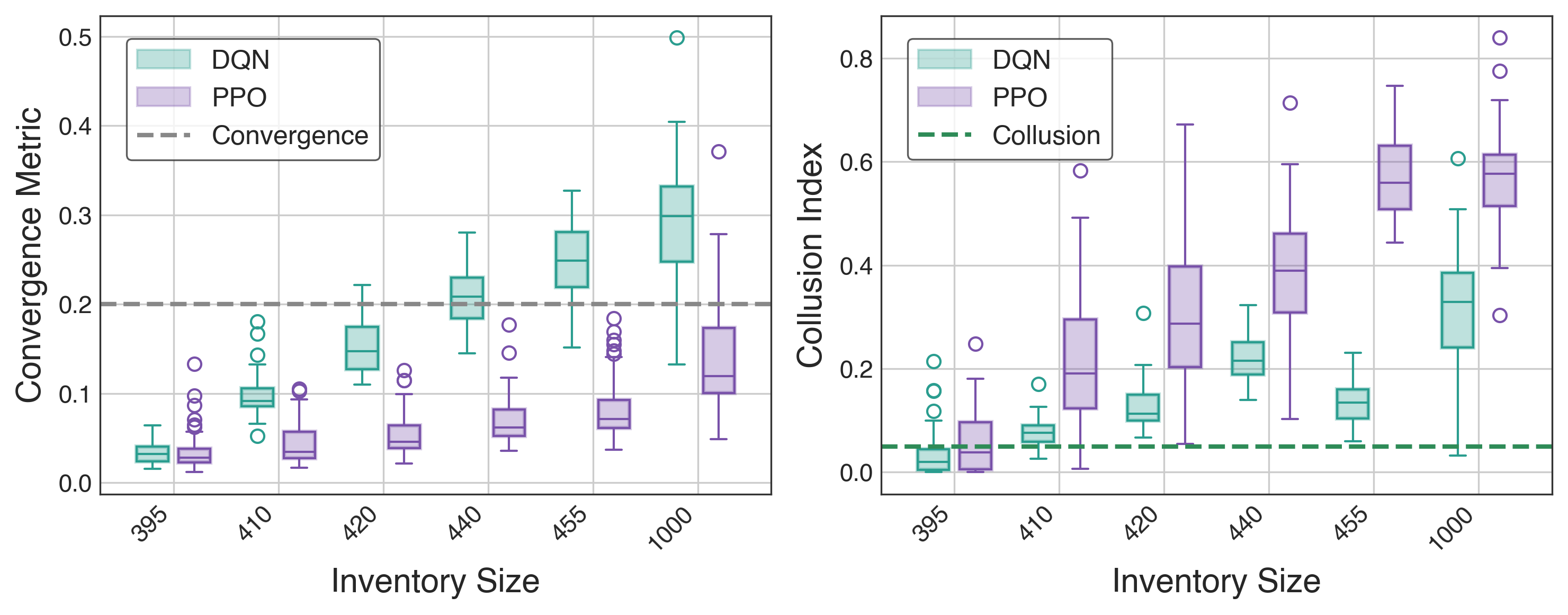}
        \vspace{-5pt}
        \caption{}
        \label{fig:DQN_PPO_boxplots:c}
    \end{subfigure}
    \begin{subfigure}[t]{\linewidth}
        \centering
        \includegraphics[width=0.9\linewidth,trim={0 10pt 0 0},clip]{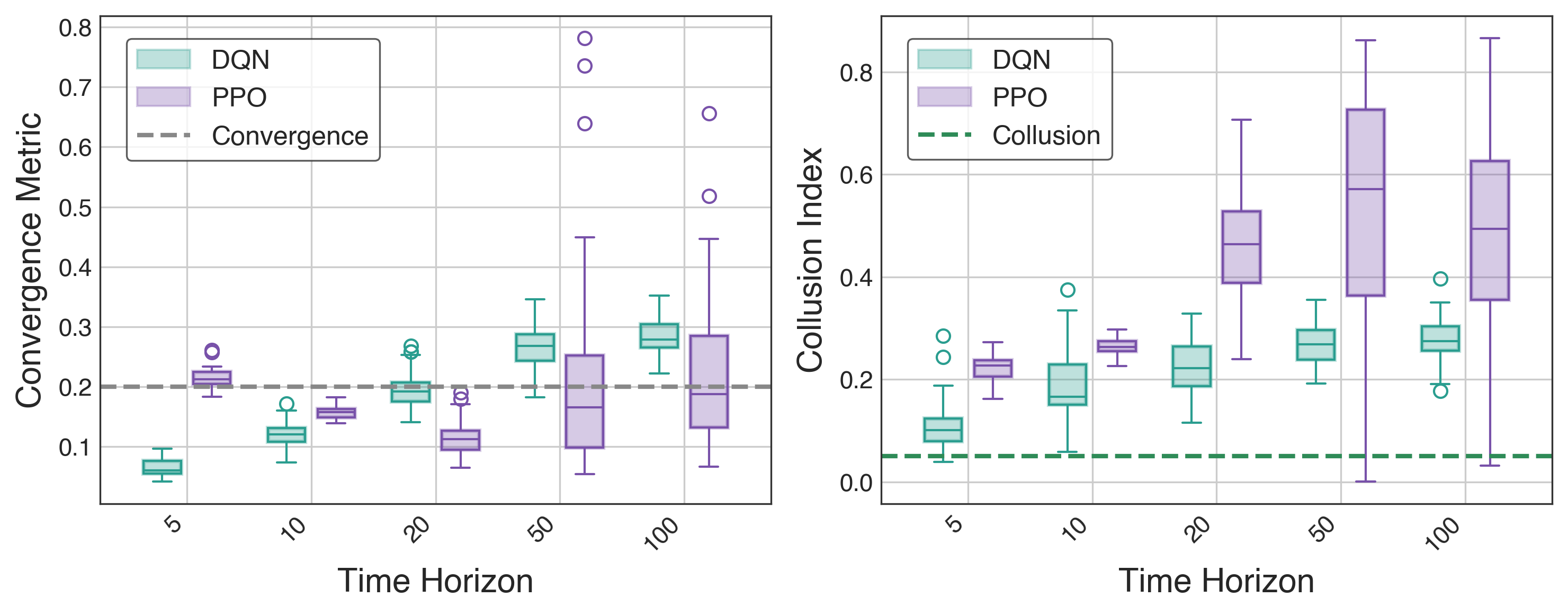}
        \vspace{-5pt}
        \caption{}
        \label{fig:DQN_PPO_boxplots:d}
    \end{subfigure}
    \label{fig:DQN_PPO_boxplots_all}
    \caption{Convergence and collusion metrics for DQN and PPO training runs with varied inventory sizes (a), and episode time horizons (b). Initial inventory size is the value shown, times the time horizon $\timehorizon$. Longer episodes show less reliable convergence, higher potential collusion due to more effective punishment strategies.}
\end{figure}

\looseness=-1
We compare metrics among different initial inventory sizes in \Cref{fig:DQN_PPO_boxplots:c}. Inventory sizes shown are per-timestep; a value of $440$ represents a total inventory size of $440 \cdot T$, which we use for the other results. Smaller inventories show better convergence and more competitive behavior for both PPO and DQN. This has geometric intuition (cf. \Cref{appendix:geometric_intuition}): visualize each agent's reward landscape as a surface over the grid of both agents' prices. Each agent tries to climb toward their peak on the side of the grid's diagonal where they undercut their opponent. Steps toward their peak along their axis harm their opponent. To achieve collusion, agents must jointly climb the ridge along the diagonal of the grid where their landscapes intersect. The closer the two agent's peaks are to the monopolistic optimum on the diagonal, the smaller their incentive to deviate and the smaller the negative impact on their opponent from deviation, easing cooperation. Decreasing inventory capacities reduces the range of prices that agents are incentivized to use as the Nash equilibrium price approaches the monopolistic price. In this ``zoomed in'' part of the price grid, the peaks now appear further away from each other, making the coordination problem harder. 

\looseness=-1
\Cref{fig:DQN_PPO_boxplots:d} shows the effect of changing episode lengths. 
As conjectured in \Cref{sec:collusion_comparison}, longer episodes increase collusion tendencies for both types of learners by providing more opportunities to punish deviations. While PPO's convergence is unaffected, DQN's convergence suffers. This is expected, as DQN generally scales worse to larger state spaces than PPO. It relies on accurately estimating the expected reward for each state-action pair and sufficiently exploring the state space, which becomes harder as that space grows.

We identified additional hyperparameters affecting collusion, such as PPO's number of training epochs (higher increases collusion) and DQN's buffer size (larger increases collusion), shown in \Cref{sec:appendix_more_hparams}. %
It is possible to hinder collusion by introducing instability in learning targets, e.g., by filling DQN's buffer or PPO's rollouts with experiences gathered from `parallel environments'. This parallelization is commonly done to increase training speed on accelerator hardware, but has a concrete impact in this model. 
We demonstrate this with PPO in \Cref{appendix:competitive_PPO}.

\section{Conclusion}
\label{sec:conclusion}
We formulate price competition between producers as an episodic Markov game motivated by Airline Revenue Management (ARM) and facilitating the analysis of tacit collusion within a finite time horizon and inventory-constrained markets.
We propose numerical methods to find competitive and collusive solutions in our model due to the lack of analytical solutions and define a collusion metric based on the total profit achieved in a full episode.
Our analysis shows that collusion consistently emerges between independent DQN and PPO algorithms after a brief period of competition and that trained agents quickly revert back to collusive prices after a forced deviation.
The proven collusive potential of RL agents in our setting covering many real markets reinforces the call for the development of mitigation strategies and regulatory efforts~\citep{calvanoProtectingConsumersCollusive2020}.

We see our work as a first step toward understanding pricing competition in markets like airline tickets, hotels, and perishable goods with future research directions in extending our Markov Game model to domain specifics.
Additionally, we see a need to consider multi-agent specific algorithms, e.g., opponent-shaping agents~\citep{soulyLeadingPackNplayer2023}, that could establish stronger collusion or even exploit market participants, significantly harming social welfare.

\clearpage

\subsection*{Acknowledgements}
Barna Pásztor is supported by an ETH AI Center doctoral fellowship.

\printbibliography

\clearpage
\appendix
\onecolumn

\section{Numerical solution strategy for Nash \& monopolistic prices}\label{appendix:gnep-solving}
\looseness=-1
To solve the GNEP for competitive equilibrium prices, we use a Gauss-Seidel-type iterative method \citep{facchineiGeneralizedNashEquilibrium2007}. We start with an initial price vector guess and proceed through a loop where each iteration updates each agent's price by solving their subproblem. For agent $i$ at iteration $k$, it uses the fixed opponent prices from the latest estimate. The process repeats until convergence to $\price^\ast$.
Each agent's subproblem is a mixed-integer, nonlinear optimization problem (MINLP), with neither convex objectives nor constraints. We use \emph{Bonmin}, a local solver capable of handling larger instances at the risk of missing global optima. We mitigate this by initiating the solver from multiple different starting points.
For the collusive optimum, we simulate a scenario where one agent sells $\numagents$ items, aiming to maximize the total episodic revenue under $\numagents$ inventory constraints. This problem is again a non-convex MINLP. Our implementation uses the open-source COIN-OR solvers via Pyomo in Python.

\section{Parameters used for training, environment, DQN and PPO}\label{appendix:params_used}
\subsection{Environment and agent parameters}\label{appendix:env_agent_params}
All our experiments use identical parameter values for both agents.

\begin{table}[h]
\centering
\begin{tabular}{@{}lr@{}}
\toprule
\textbf{Parameter}              & \textbf{Value} \\ \midrule
Product quality $\quality_i$    & \num{2} \\
Outside good quality $\quality_0$ & \num{0} \\
Marginal cost $\cost_i$         & \num{1} \\
Horizontal differentiation $\horizontaldiff$    & \num{0.25} \\
Time horizon $\timehorizon$     & \num{20} \\
Demand scaling factor $\demandscalingfactor$    & \num{1000} \\
Inventory capacity $\capacity_i$              & $440 * \timehorizon$ \\
Nash price $\price^N$ (unconstrained)           & \num{1.471} \\
Nash price $\price^N$ (constrained)             & \num{1.693} \\
Monopolistic price $\price^M$   & \num{1.925} \\
Number of prices in interval    & \num{15} \\
Price interval parameter $\xi$  & \num{0.2} \\
\bottomrule
\end{tabular}
\caption{Environment and agent parameters}
\label{table:env_params}
\end{table}

\subsection{DQN and PPO hyperparameters}\label{appendix:dqn_ppo_params}
We used the same neural network architecture for both DQN and PPO, of $2$ hidden layers with $64$ neurons each. These hyperparameters were found by starting with generally accepted values from reference implementations and refined by doing grid searches over up to three parameters at a time.

DQN's epsilon-greedy strategy's epsilon parameter is annealed via an exponential decay from an initial value of $\epsilon_{\max} = 1$ to $\epsilon_{\text{min}}=0.015$ at the end of the training run. \\
At training episode $e \in \{0, \ldots, \numepisodes\}$, epsilon's value is $\epsilon_{\max} * \bigl(\frac{\epsilon_{\text{max}}}{\epsilon_{\text{min}}}\bigr)^{e/\numepisodes}$.
\begin{table}[h]
\centering
\begin{tabular}{@{}lr@{}}
\toprule
\textbf{Parameter}              & \textbf{Value} \\ \midrule
Training episodes $\numepisodes$    & \num{50000} \\
Learning rate                   & \num{0.001} \\
Adam epsilon                    & \num{0.001} \\
Epsilon-greedy (annealed) $\epsilon_{\text{min}}$   & \num{0.015} \\
Replay buffer size              & \num{200000} \\
Replay buffer batch size        & \num{64} \\
Gradient norm clipping          & \num{25} \\
Initial episodes without training                   & \num{5000} \\
Train agent every ... episodes                  & \num{4} \\
Target network update every ... episodes            & \num{200} \\
Network layer sizes             & $[64, 64]$ \\
\bottomrule
\end{tabular}
\caption{DQN hyperparameters}
\label{table:dqn_params}
\end{table}

PPO's entropy coefficient parameter is annealed via an exponential decay from an initial value of $\text{ent}_{\text{max}} = 0.03$ to $\text{ent}_{\text{min}} = 0.0001$ at $75\%$ of the training run (and is clipped to $\text{ent}_{\text{min}}$ afterwards). \\
At training episode $e \in \{0, \ldots, \numepisodes\}$, the coefficient's value is $\text{ent}_{\text{max}} * \bigl(\frac{\text{ent}_{\text{min}}}{\text{ent}_{\text{max}}}\bigr)^{e/0.75\numepisodes}$.

\begin{table}[h]
\centering
\begin{tabular}{@{}lr@{}}
\toprule
\textbf{Parameter}              & \textbf{Value} \\ \midrule
Training episodes $\numepisodes$    & \num{1000} \\
Learning rate                   & \num{2.5e-4} \\
Adam epsilon                    & \num{1e-5} \\
Number of minibatches           & \num{10} \\
Number of training epochs       & \num{20} \\
GAE-lambda                      & \num{0.95} \\
Value coefficient (with clipping)   & \num{0.5} \\
Gradient norm clipping          & \num{0.5} \\
Network layer sizes             & $[64, 64]$ \\
\bottomrule
\end{tabular}
\caption{PPO hyperparameters}
\label{table:ppo_params}
\end{table}

\section{Literature review}\label{appendix:litreview}
\paragraph{Examples and description of tacit collusion} Firms across various sectors, from insurance to flight tickets, employ \emph{algorithmic pricing} to maximize revenue by leveraging data on market conditions, customer profiles, and other factors. These algorithms' growing complexity raises challenges for maintaining fair competition and detect firms that \emph{tacitly collude}, ones which jointly set \emph{supra-competitive} prices (i.e., above the competitive level) or limit production \emph{without explicit agreements or communication}. Recently, evidence has emerged that companies are already using algorithmic pricing to inflate prices market-wide at the cost of consumers. 
For instance,~\citet{assadAlgorithmicPricingCompetition2024} showed that German fuel retailer margins increased by 38\% following the widespread adoption of algorithmic pricing. Other examples are found in setting credit card interest rates~\citep{ausubel1991failure} and consumer goods markets~\citep{genesove2001rules}. 

\paragraph{Legal developments around algorithmic collusion} Current anti-collusion policies mainly address explicit agreements, making tacit collusion inferred from company behaviors rather than evidence of an agreement, more elusive %
to prove. There is growing concern among regulators~\citep{ohlhausenShouldWeFear2017, bundeskartellamtAlgorithmsCompetition2019, directorate-generalforcompetitioneuropeancommissionCompetitionPolicyDigital2019} and researchers~\citep{harringtonDevelopingCompetitionLaw2018, benekeRemediesAlgorithmicTacit2021, breroLearningMitigateAI2022} that AI-based pricing algorithms might evade competition laws by colluding tacitly, without direct communication or explicit instruction during learning. This highlights the need for better strategies to prevent collusion or mitigate its negative effects on the market. 

\paragraph{Reinforcement learning (RL) background}
\emph{Reinforcement learning}~\citep{sutton2018reinforcement} is an advanced segment of machine learning where agents learn to make sequential decisions by interacting with an environment. 
Unlike traditional machine learning methods which rely on static datasets, RL emphasizes the development of autonomous agents that improve their behavior through trial-and-error, learning from their own experiences. This approach enables agents to understand complex patterns and make optimized decisions in scenarios with uncertain or shifting underlying dynamics. 
\emph{Multi-agent} RL extends this concept to scenarios involving multiple decision-makers, each optimizing their strategies while interacting with others and the environment~\citep{bucsoniu2010multi}. In MARL settings, agents can be incentivized to behave competitively, as seen in zero-sum games like Go~\citep{silverMasteringGameGo2017,silver2018general}, cooperatively, like in autonomous vehicle coordination~\citep{dinnewethMultiagentReinforcementLearning2022} or a mix of the two that includes our problem, i.e., markets and pricing games. MARL, while posing challenges such as \emph{non-stationarity} and \emph{scalability}, enables agents to adapt to and influence competitors' strategies, facilitating tacit collusion.

\paragraph{Collusion \& regulation in airline revenue management (ARM)}
Originally a strictly regulated sector with price controls, ARM was deregulated in 1978 in the US and Europe, leading to a competitive landscape of private carriers whose pricing strategies are subject only to general laws against anti-competitive behavior~\citep{TFEU2012}(Art. 101-109). However, this deregulation has caused market consolidation, prompting regulatory responses to protect competition~\citep{EUreg2019}. Even prior to algorithmic pricing, regulators have identified pricing behaviors suggestive of tacit collusion~\citep{borenstein1994competition}, underscoring the challenge of distinguishing between collusive behavior and independent but parallel responses to market conditions. 

\paragraph{Background on the field of revenue management (RM)}
Each of the agents that we model is individually maximizing their revenue, relating our work to the field of \emph{revenue management (RM)}~\citep{talluriTheoryPracticeRevenue2004}. As a competitive market with slim net margins, airlines are increasingly turning to \emph{dynamic pricing}~\citep{koenigsbergEasyJetAirlines2004} beyond traditional \emph{quantity-based} and \emph{price-based} RM, replacing the hugely popular expected marginal seat revenue (EMSR) models~\citep{Belobaba1987}.
Our problem falls into the price-based RM category, even though we do model aspects of capacity management with our inventory constraints. In quantity-based RM, agents decide on a production quantity with the price for their good being the result of a market-wide fixed function of that decision, and models often impose no limit on the offered quantity. In our model, agents decide their price, and demand results from a market-wide function. Our aim is that agents learn to predict the impact of their pricing choices on the demand and thus sold quantity, in order to optimally use their constrained inventory.

\paragraph{Learning in general RM}
In recent years, reinforcement learning agents have seen increased use in revenue management outside of the airline context. Examples include learning both pricing and production quantity strategies in a market with perishable goods~\citep{Wang2021}, producing a pricing policy by learning demand~\citep{ranaRealtimeDynamicPricing2014, ranaDynamicPricingPolicies2015} and analyzing the performance of different popular single-agent RL in various market settings~\citep{kastiusDynamicPricingCompetition2022a} (here Q-learning and Actor-Critic). The use of largely uninterpretable learned choice or pricing models introduces new challenges, such as deriving economic figures like the elasticity of demand with respect to price~\citep{acuna-agostPriceElasticityEstimation2023}.

\paragraph{Learning in ARM}
While early work used e.g. heuristically solved linear programming formulations~\citep{brontColumnGenerationAlgorithm2009} or custom learning procedures~\citep{vanryzinRevenueManagementForecasting2000, bertsimasSimulationBasedBookingLimits2005}, recent studies have explored single-agent reinforcement learning in ARM to learn optimal pricing~\citep{Razzaghi2022}. These model the problem as a single-agent Markov decision problem (MDP)~\citep{gosaviReinforcementLearningApproach2002, lawheadBoundedActorCritic2019} and consider realistic features like cancellations and overbooking~\citep{Shihab2022}. The application of \emph{deep reinforcement learning (deep-RL)}~\citep{mnihHumanlevelControlDeep2015} is growing in this complex market~\citep{Bondoux2020, Alamdari2021}, but these models often overlook the multi-agent nature of the airline market. We model the market as a multi-agent system with individual multi-agent learners, a critical yet unexplored aspect in current research~\citep{Razzaghi2022}.

\section{Proof of Lemma 1}\label{appendix:proof-lemma1}
\begin{proof}
    Let us introduce some terminology first. 

    \begin{definition}
        Fix an agent $i$ with policy $\policy_i$ or price vector $p^{(i)}$, and fix opponent policies $\policy^{(-i)}$ or prices $\price^{(-i)}$. 
    \begin{itemize}
        \item A \emph{useful deviation} is a policy $\policy'_i$ or price vector $\price^{(i)'}$ that strictly increases $i$'s revenue over the whole episode compared to playing $\policy_i$ or $\price^{(i)}$. We use this term in both the Markov game and SMG. 
        \item We call a price vector $\price^{(i)} = (\price_{i,1},\ldots,\price_{i,T})$ \emph{feasible in the GNEP} if it fulfills the inventory constraint of $i$'s revenue maximization problem in \Cref{eq:gnep}, and \emph{infeasible in the GNEP} if it does not.
        \item We call a policy $\policy_i$ \emph{simple}, if at each time $t$, it outputs the same value for all states $\agentstate_t$, i.e. $\forall t\ \forall \agentstate_t: \policy_i(\agentstate_t) \equiv \text{const}_t$.  
    \end{itemize}
    \end{definition}

    Intuitively, we construct a set of simple policies where each agent always plays their GNEP solution, no matter the state, and show that this set of policies is a Nash equilibrium.

    First, observe that those simple policies result in the same set of price vectors $p^\ast$ in every evolution of the Markov game. In particular, fixing opponent strategies $\policy^{(-i)\ast}$ results in agent $i$ facing the same fixed opponent price vectors $\price^{(-i)\ast}$ (from the GNEP solution) in every evolution of the Markov game. Therefore, to prove that $\policy^\ast$ is a Nash equilibrium in the Markov game it is enough to prove that for any agent $i$ and fixed opponent price vectors $\price^{(-i)\ast}$, there does not exist a useful deviation price vector $\price^{(i)'} \neq \price^{(i)}$. If a useful deviation policy $\policy'_i$ existed for $i$, in at least one timestep $t$ it would have to pick a price $\price'_{i,t} \neq \price_{i,t}$, so by ruling out a useful price vector deviation we also rule out a useful policy deviation.

    \textbf{Claim:} Let $\price^{(-i)}$ be fixed opponent price vectors. Given any price vector $p^{(i)}$ for agent $i$, there always exists a price vector $\Bar{\price}^{(i)}$ that is feasible in the GNEP and such that playing $\Bar{\price}^{(i)}$ results in revenue for $i$ that is as great as or greater than that from playing $p^{(i)}$. 
    
    Given opponent prices $\price^{(-i)\ast}$, if a useful deviation $\price^{(i)'} \neq \price^{(i)\ast}$ exists for agent $i$, it must be infeasible in the GNEP (otherwise $\price^{(i)\ast}$ wouldn't be a revenue-maximizing solution to agent $i$'s GNEP's subproblem). However, since the claim implies that we could construct a $\Bar{\price}^{(i)}$ that is feasible in the GNEP and has equivalent revenue for $i$ as the infeasible $\price^{(i)'}$, it would be a useful deviation for agent $i$ in the SMG to play $\Bar{\price}^{(i)}$ given $\price^{(-i)\ast}$, contradicting the assumption that $p^\ast$ is a NE.

    \textbf{Proof of Claim:} Let opponent prices be fixed $\price^{(-i)}$. Let $\price^{(i)}$ a price vector in the Markov game that's infeasible in the GNEP (otherwise we're trivially done). Let $i$'s inventory at $t$ be $\inventory_t$. Let $\hat{t} \in \{1,\ldots,\timehorizon\}$ be the \emph{sell-out time}, i.e., the last timestep in which $i$ has nonzero inventory, meaning $\hat{t} := \max\{t \in \{1, \ldots, \timehorizon\} |\inventory_{\hat{t}} > 0\}$ such that $\inventory_{\hat{t}}=0$ and $\forall t>\hat{t}: \inventory_t = 0$. Let $\demand(\price_{i,t}, \price_{(-i),t}) := \lfloor \demandscalingfactor \demand_{i,t} \rfloor$ be the scaled, truncated MNL demand of agent $i$ at time $t$ given price vector $\price$, which is a decreasing function in $\price_{i,t}$.
    
    Define 
    \begin{align*}
        \Bar{\price}_{i,\hat{t}} &:= \sup\{q\ |\ \demand(q, \price_{(-i),\hat{t}}) = \inventory_{\hat{t}}\} \\
        \Bar{\price}_{i,t} &\in \{q\ |\ \demand(q, \price_{(-i),t}) = 0\}\quad \forall t > \hat{t}.
    \end{align*}
    Then, let $\Bar{\price}^{(i)} := (\price_{i,1}, \ldots, \price_{i,\hat{t}-1}, \Bar{\price}_{i,\hat{t}}, \Bar{\price}_{i,\hat{t}+1}, \ldots, \Bar{\price}_{i,T})$. 
    
    Given the other agents' fixed price vectors $\price^{(-i)}$, the vector $\Bar{\price}^{(i)}$ is feasible in the GNEP. To see this, consider that every price vector has a sell-out time $\hat{t}$. At any point in time before $\hat{t}$, the accumulated demand up until that time is lower than inventory, otherwise $\hat{t}$ wouldn't actually be the sell-out time. The GNEP's feasibility constraint is only violated if at $\hat{t}$, demand is larger than remaining inventory $\inventory_{\hat{t}}$, or if at any $t > \hat{t}$, demand is larger than $0$. The construction of $\Bar{\price}^{(i)}$ ensures that it has the same sell-out time $\hat{t}$, and the construction of $\Bar{\price}_{i,t}$ for $t\geq \hat{t}$ ensures that demand at $\hat{t}$ matches inventory left, and that demand at $t>\hat{t}$ is zero, meaning that $\Bar{\price}^{(i)}$ cannot violate the feasibility constraint.
    
    Now we just need to prove that given fixed opponent prices $\price^{(-i)}$, agent $i$'s reward in the Markov game when playing $\Bar{\price}^{(i)}$ is as great as or greater than their reward when playing $\price^{(i)}$. Their reward when playing $\price^{(i)}$ is given by 

    \begin{align*}
        &\Sigma_{t=1}^{\hat{t}-1} (\price_{i,t} - \cost) \min\left(\demand(\price_{i,t}, \price_{(-i),t}), \inventory_t\right) \\
        &+ (\price_{i,\hat{t}} - \cost)\min\left(\demand(\price_{i,\hat{t}}, \price_{(-i),\hat{t}}), \inventory_{\hat{t}}  \right) \\
        &+ \Sigma_{t=\hat{t}+1}^{T} (\price_{i,t} - \cost) \min\left(\demand(\price_{i,t}, \price_{(-i),t}), \inventory_t\right)
    \end{align*}
    
    We now replace $\price^{(i)}$ with $\Bar{\price}^{(i)}$ and compare each term. 
    
    In the \emph{first term}, as we know that for $t < \hat{t}$ $i$'s demand is always lower than their inventory by definition of $\hat{t}$, the term reduces to 
    \begin{equation*}
        \Sigma_{t=1}^{\hat{t}-1} (\price_{i,t} - \cost) \demand(\price_{i,t}, \price_{(-i),t}).
    \end{equation*}
    Since $\price_t = \Bar{\price}_t$, we see that the first revenue term's value stays equal:

    \begin{align*}
        &\Sigma_{t=1}^{\hat{t}-1} (\price_{i,t} - \cost) \min\left(\demand(\price_{i,t}, \price_{(-i),t}), \inventory_t\right) \\
        &= \Sigma_{t=1}^{\hat{t}-1} (\price_{i,t} - \cost) \demand(\price_{i,t}, \price_{(-i),t}) \\
        &= \Sigma_{t=1}^{\hat{t}-1} (\Bar{\price}_{i,t} - \cost) \demand(\Bar{\price}_{i,t}, \price_{(-i),t}).
    \end{align*}

    In the \emph{second term}, by definition of $\hat{t}$, we know that 
    
    $$\min\left(\demand(\price_{i,\hat{t}}, \price_{(-i),\hat{t}}), \inventory_{\hat{t}}\right) = \demand(\price_{i,\hat{t}}, \price_{(-i),\hat{t}}) = \inventory_{\hat{t}},$$
    
    thus the term reduces to 
    \begin{equation*}
        (\price_{i,\hat{t}}- \cost)\demand(\price_{i,\hat{t}}, \price_{(-i),\hat{t}}).
    \end{equation*}
    Since $\demand(\price_{i,\hat{t}}, \price_{(-i),\hat{t}}) \geq \inventory_{\hat{t}}$, and by construction $\demand(\Bar{\price}_{i,\hat{t}}, \price_{(-i),\hat{t}}) = \inventory_{\hat{t}}$, and $d(\cdot, \price_{(-i),\hat{t}})$ decreasing, we get $\Bar{\price}_{i,\hat{t}} \geq \price_{i,\hat{t}}$. We also know that $i$ will always choose a price $\geq \cost$ to ensure non-negative revenue. Thus, we see that the second revenue term's value can only increase:

    \begin{align*}
        &(\price_{i,\hat{t}} - \cost) \min\left(\demand(\price_{i,\hat{t}}, \price_{(-i),\hat{t}}), \inventory_{\hat{t}} \right) \\
        &= (\price_{i,\hat{t}} - \cost)\demand(\price_{i,\hat{t}}, \price_{(-i),\hat{t}}) \\
        &\leq (\Bar{\price}_{i,\hat{t}} - \cost)\demand(\Bar{\price}_{i,\hat{t}}, \price_{(-i),\hat{t}}).
    \end{align*}

    In the \emph{third term}, by definition of $\hat{t}$, we know that $\forall t > \hat{t}: \inventory_t = 0$, and since by construction of $\Bar{\price}^{(i)}$ we also know that $\forall t > \hat{t}: \demand(\Bar{\price}_{i,t}, \price_{(-i),t}) = 0$, we see that the term's value remains zero:
    
    \begin{align*}
        &\Sigma_{t=\hat{t}+1}^{T} (\price_{i,t} - \cost) \min\left(\demand(\price_{i,t}, \price_{(-i),t}), \inventory_t\right) \\
        &= \Sigma_{t=\hat{t}+1}^{T} (\Bar{\price}_{i,t} - \cost) \demand(\Bar{\price}_{i,t}, \price_{(-i),t}) \\
        &= 0
    \end{align*}

    Putting all three terms together, agent $i$'s revenue from playing $\Bar{\price}^{(i)}$ is as great as, or greater than that from playing $\price^{(i)}$.
\end{proof}

\section{Supplementary experiments}\label{appdx:additional_experiments}
\subsection{Comparison of means}\label{appdx:means_comparison}
\begin{figure}[htbp]
    \centering
    \includegraphics[width=\linewidth]{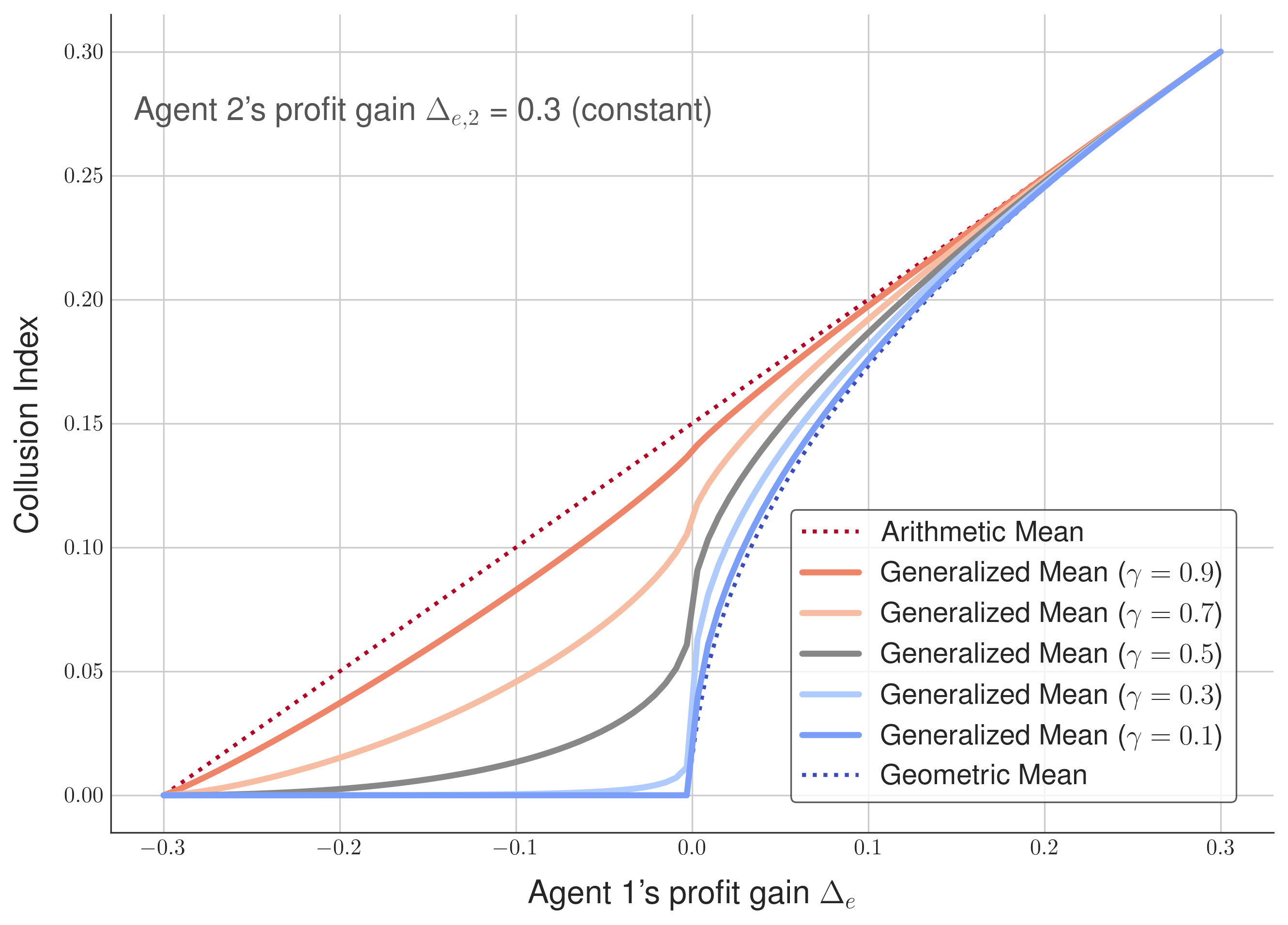}
    \caption{Showing how the generalized mean interpolates between the arithmetic and geometric means for choices of parameter $\gamma \in [0,1]$, in the context of the collusion index measure.}
    \label{fig:means_comparison}
\end{figure}

\Cref{fig:means_comparison} compares the arithmetic, geometric and generalized means. We vary the generalized mean's parameter $\gamma$ between $0$ and $1$, showing that it interpolates the arithmetic and generalized means, providing a balance between the former's ability to deal with negative values, and the latter's ability to weigh outcomes with supra-competitive total profits, but a disparity in profit gain between agents, as less collusive than symmetrical ones.

\subsection{PPO behavior analysis}\label{sec:PPO_analysis}
\begin{figure}[htbp]
    \centering
    \includegraphics[width=\linewidth]{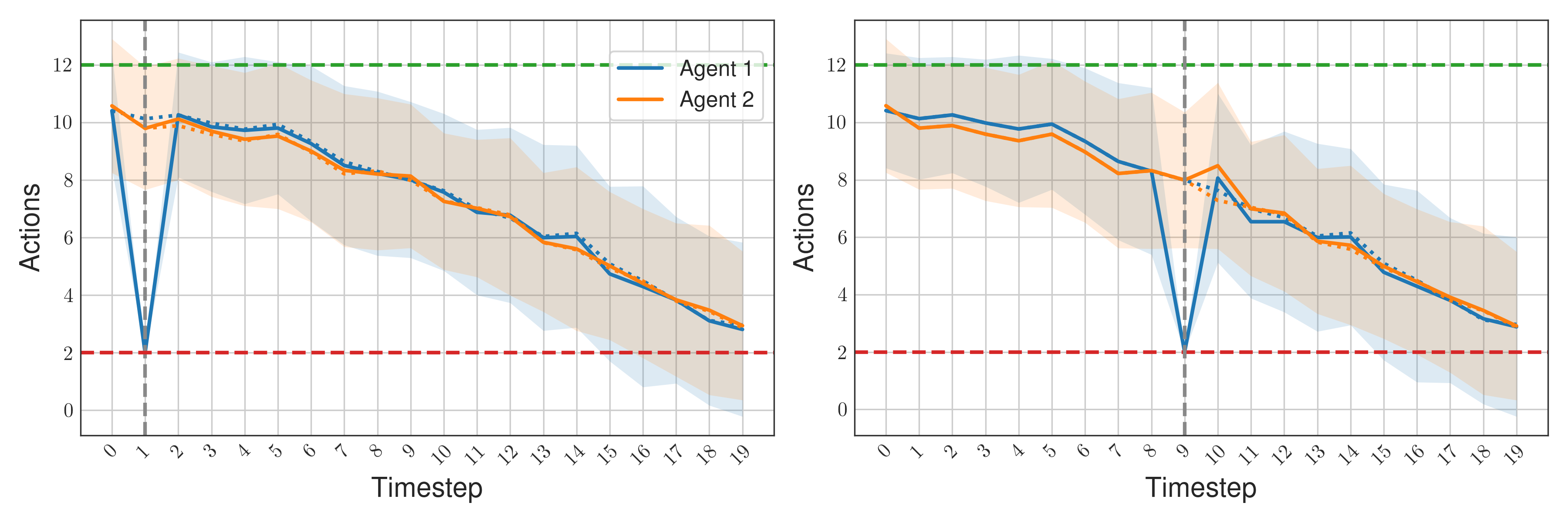}
    \caption{Behavior of two PPO agents during an episode after forcing one agent to deviate at time $t=1$ and $t=9$ respectively. Dotted lines indicate evolution without deviation. Deviations provoke a competitive reaction, with both agents quickly returning to collusion.}
    \label{fig:PPO_forced_dev}
\end{figure}

\begin{figure}[htb]
    \centering
    \includegraphics[width=\linewidth]{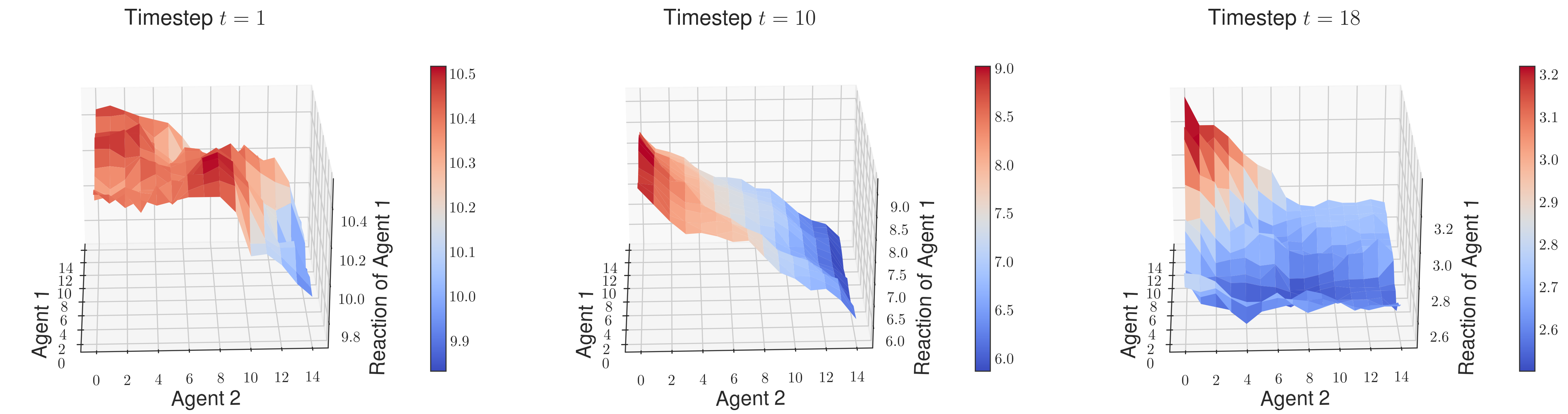}
    \caption{``PPO response surface''. The surfaces show a PPO agent 1's learned response to a state given by both agent's prices (x- and y-axes), timestep and symmetric remaining inventory level.}
    \label{fig:PPO_reaction_surface}
\end{figure}

Like in \Cref{sec:exp_strategy_analysis}, we analyze PPO's learned strategies from its behavior during an eval episode in \Cref{fig:PPO_forced_dev} and from its response surfaces. The evaluation episode shows that PPO has learned to collude over multiple episodes, with both agents starting off highly collusive and gradually undercutting each other, ending up at the collusive level at the end of the episode. This contrasts DQN's tendency to rise in collusion during the episode, before defecting toward the end. PPO does not seem to punish collusion strongly. Its reaction surface (\Cref{fig:PPO_reaction_surface}) suggests that it instead relies on a mutual understanding of collusion as a slow price war, undercutting if the opponent prices higher but preferring to reset the price after an opponent's deviation.

\subsection{Choice of price-action grid}\label{sec:appendix_price_action}
\begin{figure}[h!]
    \centering
    \includegraphics[width=\linewidth]{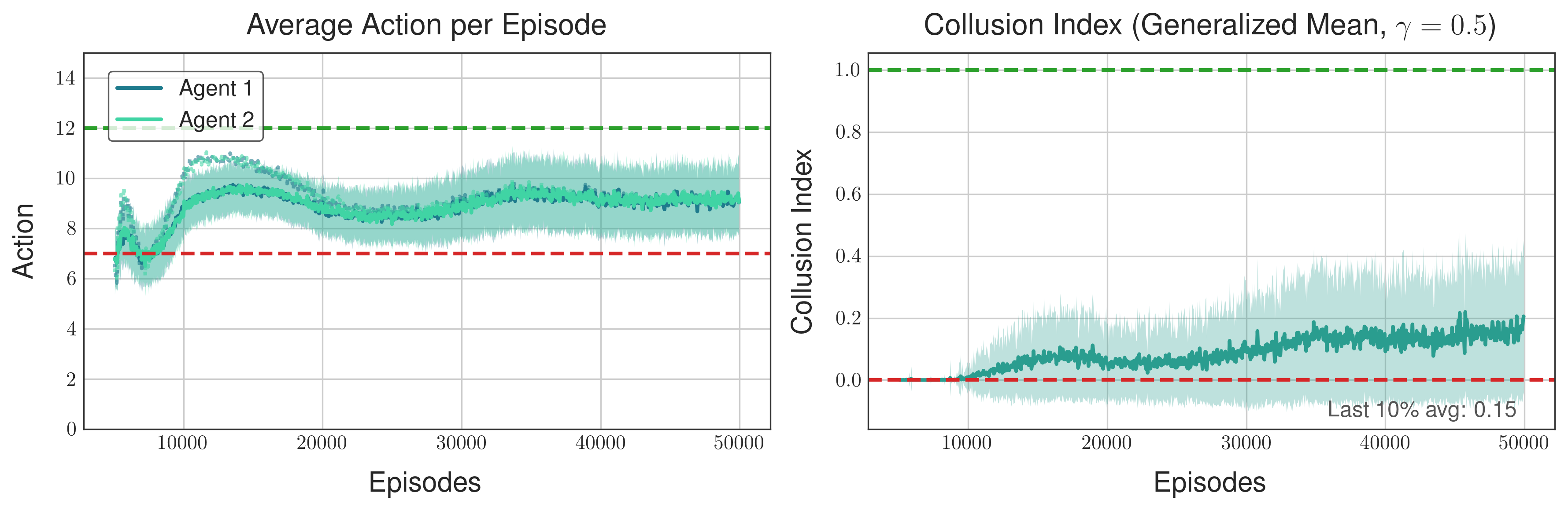}
    \caption{Choice of price-action grid: Training run evolution of two DQN learners using a price grid spanning the interval between the unconstrained Nash and monopolistic prices rather than the narrower constrained ones. Agents show the same pattern of learning competition and collusion.}
    \label{fig:full_price_grid}
\end{figure}

In the constrained setting, we define the available actions to be a discretized grid in the interval between (and extending slightly beyond) the \emph{constrained} Nash equilibrium and monopolistic optimum prices. This interval is narrower than in the unconstrained case, as the Nash equilibrium price increases, while the monopolistic price level stays the same. In an episodic setting, agents are effectively unconstrained at the beginning of an episode, so by doing this we are restricting some of their ability to strategize. However, as \Cref{fig:full_price_grid} shows, agents quickly learn to only price between the constrained competitive and collusive interval, suggesting that restricting the price grid does not cut off a relevant part of the strategy space. 

\subsection{Asymmetric inventory constraints}
\label{appendix:asymmetric_inventory}

\begin{figure}[htbp]
    \centering
    \includegraphics[width=\linewidth]{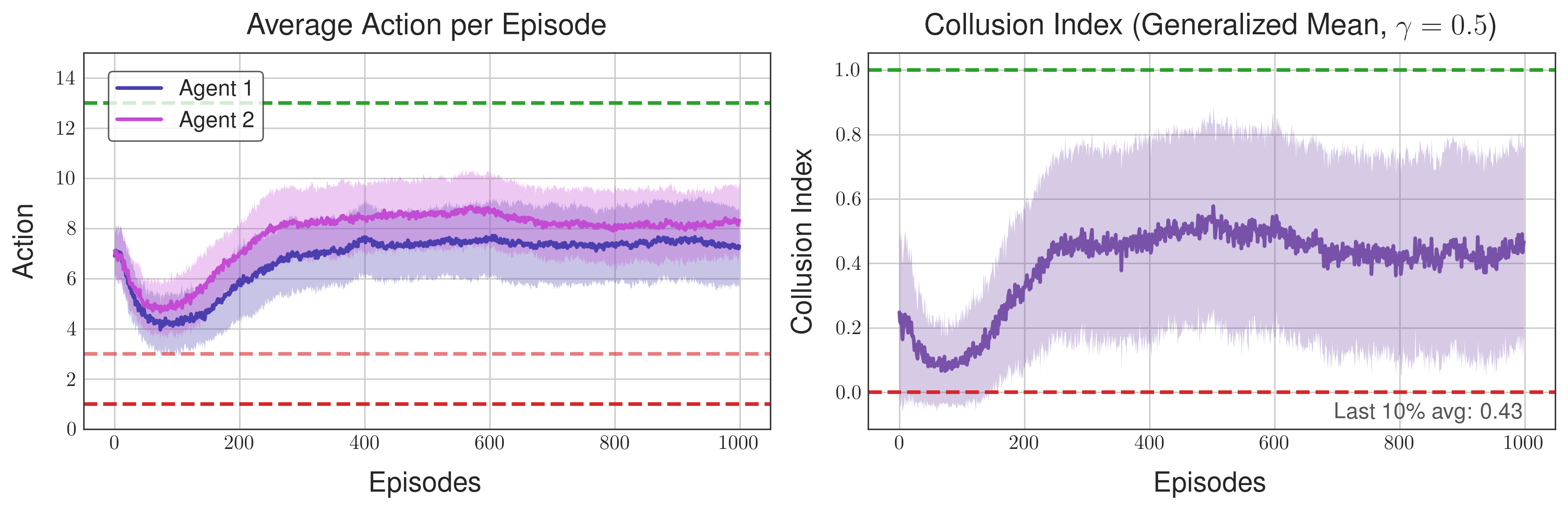}
    \caption{Asymmetric inventory constraints: Training run evolution of two PPO learners with asymmetric constraints of $440 \cdot \timehorizon$ and $400 \cdot \timehorizon$ respectively. Both agents settle at their now different collusive equilibrium price levels, but the overall collusion level remains the same.}
    \label{fig:uneven_inventory}
\end{figure}

We have assumed symmetric inventory constraints so far. Illustrated in \Cref{fig:uneven_inventory}, PPO agents with asymmetric inventory constraints of $440$ and $400$ per timestep respectively (compared to $440$ for both agents in the main paper), obtain a collusion index of 0.44 compared to the $0.43$ of the symmetric case. Both agents price collusively, although the chosen price levels now differ: the agent with the tighter inventory constraint chooses an even higher action than before. This suggests that small asymmetries do not impact the emergence of collusion.

\subsection{Observability}
\label{appendix:observability}
\begin{figure}[h!]
    \begin{subfigure}[h]{\linewidth}
        \includegraphics[width=\linewidth]{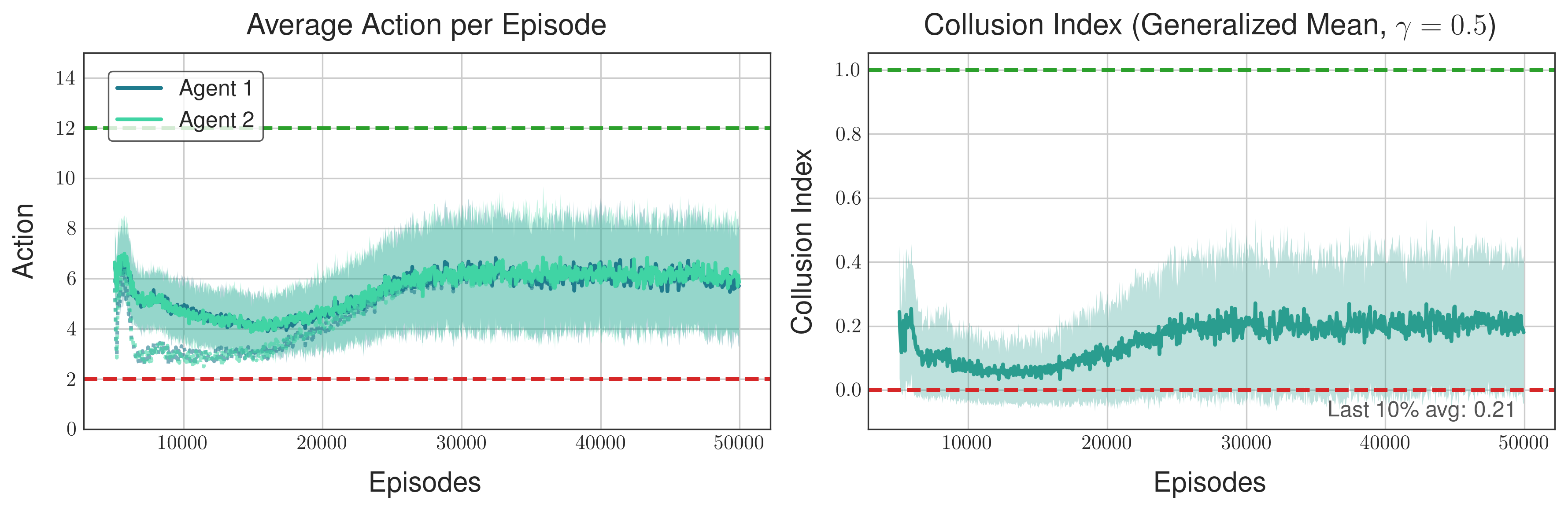}
        \label{fig:observability:no_opp_inventory}
    \end{subfigure}
    \begin{subfigure}[h]{\linewidth}
        \includegraphics[width=\linewidth]{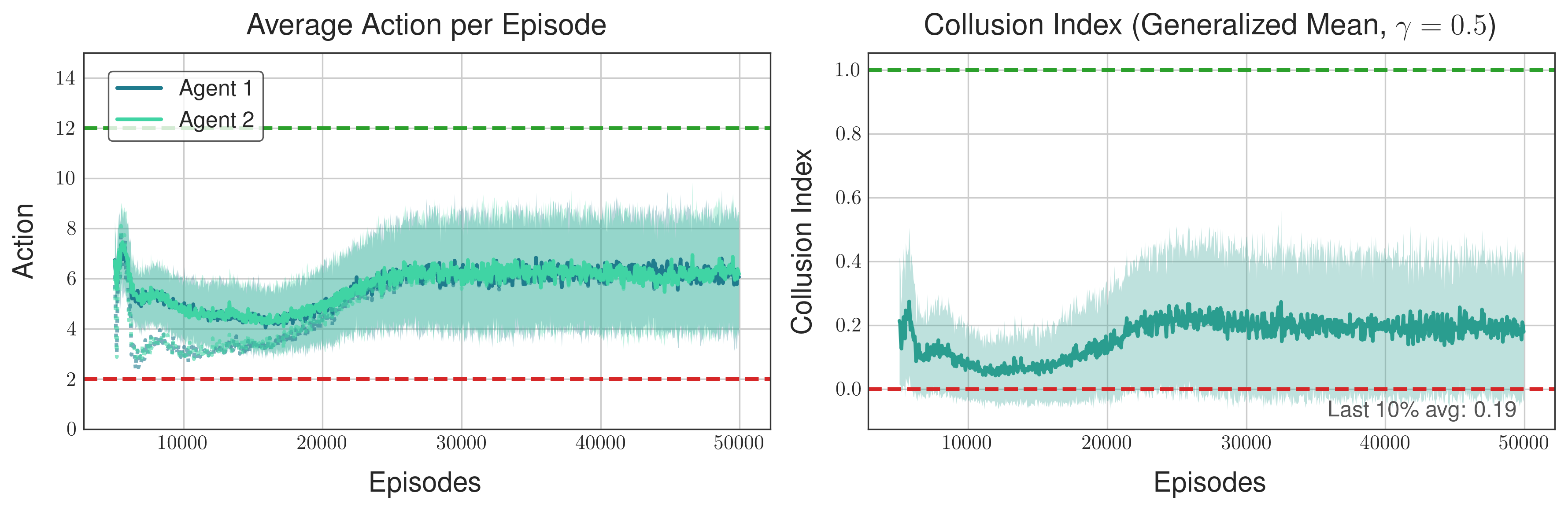}
        \label{fig:observability:no_t}
    \end{subfigure}
    \caption{Observability: Evolution of training runs of two DQN learners that are prevented from observing their opponent's inventory (top) or the current timestep (bottom). We do not see a significant difference in behavior and collusive tendency.}
    \label{fig:observability}
\end{figure}

So far, we have assumed full observability. To test that assumption's impact, we perform additional experiments. \Cref{fig:observability} shows the evolution for DQN, qualitative results are identical for PPO.

First, we remove the opponent's inventory from an agent's observation. This has little impact on DQN's performance, with the collusion index dropping from $0.43$ to $0.41$ for PPO, and from $0.23$ to $0.21$ for DQN. Second, we do not allow agents to observe time within an episode. We see only a small impact on DQN's performance. The collusion index drops to $0.36$ for PPO, and to $0.19$ for DQN.

\subsection{Geometric intuition on impact of inventory constraints on collusion}
\label{appendix:geometric_intuition}

\begin{figure}[h!]
    \centering
    \begin{subfigure}[b]{0.30\textwidth}
        \centering
        \includegraphics[width=\textwidth,trim={48pt 20pt 0pt 50pt},clip]{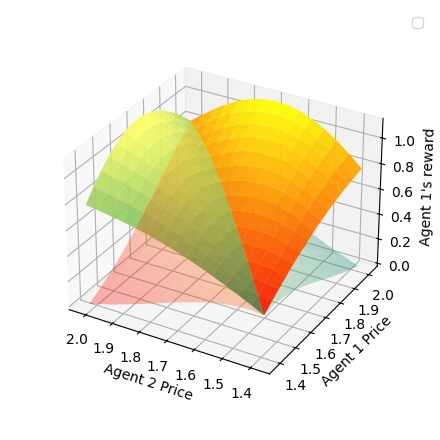}
        \caption{Unconstrained inventory}
        \label{fig:reward_surface1}
    \end{subfigure}
    \hfill
    \begin{subfigure}[b]{0.30\textwidth}
        \centering
        \includegraphics[width=\textwidth,trim={48pt 20pt 0pt 50pt},clip]{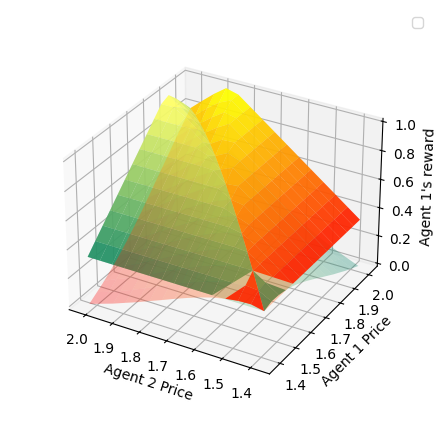}
        \caption{Light constraint}
        \label{fig:reward_surface2}
    \end{subfigure}
    \hfill
    \begin{subfigure}[b]{0.30\textwidth}
        \centering
        \includegraphics[width=\textwidth,trim={48pt 20pt 0pt 50pt},clip]{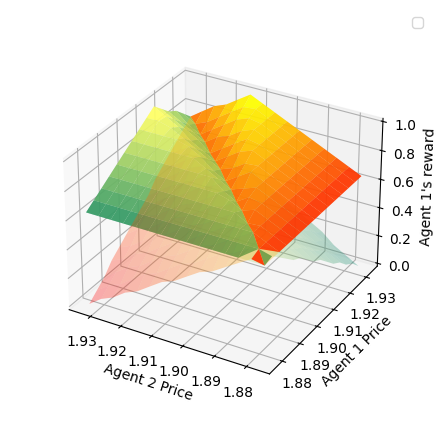}
        \caption{Strong constraint}
        \label{fig:reward_surface3}
    \end{subfigure}
    \caption{Geometric intuition on impact of inventory constraints on collusion: Normalized reward surfaces of agent 1 (green) and 2 (red) in a single period as functions of the prices both agents choose, with symmetric inventory capacities of different sizes (left to right: infinite, $470$, $380$). Introducing an inventory constraint pushes the peaks of both reward surfaces closer to each other, but tightening the constraints and thus 'zooming in' on the area near the peaks makes the peaks appear further apart, hindering collusion.}
    \label{fig:reward_surfaces}
\end{figure}

\Cref{fig:reward_surfaces} shows the normalized reward (profit) surfaces of agent 1 (green) and 2 (red) of the one-period game as functions of the prices both agents choose. We compare the surface under symmetric inventory capacities of different sizes, namely unconstrained (i.e., infinitely large inventory), lightly constrained (capacity $470$) and strongly constrained (capacity $380$). One can observe that the peak of each agent's reward surface lies on the opposite side of the diagonal through the price grid as their opponent's, namely on the side where they undercut their opponent's price and profit from capturing additional demand. Introducing an inventory constraint limits agents' profits from undercutting as they cannot capture the additional demand past their inventory limit. This pushes the peaks of both agents' reward surfaces closer to the diagonal, and each other. Further tightening the constraints while keeping the number of actions between the Nash and collusive optima equal and thus 'zooming in' on the price grid to an area near the peaks makes the peaks appear further apart. As mentioned in the main paper, we can imagine the duopoly dynamic as both learning agents trying to climb toward the peak of their respective reward surface. Collusion is achieved if agents climb the ridge along the diagonal. The closer the peaks are to each other and thus the monopolistic optimum on the diagonal, the smaller each agent's incentive to deviate and the smaller the negative impact of a deviation on their opponent, which eases cooperation. Tightening inventory constraints thus complicates the coordination problem, making collusion less likely.

\subsection{Steering PPO toward competitive behavior}\label{appendix:competitive_PPO}
\begin{figure}[htbp]
    \centering
    \begin{subfigure}[h]{\linewidth}
        \centering
        \includegraphics[width=\linewidth]{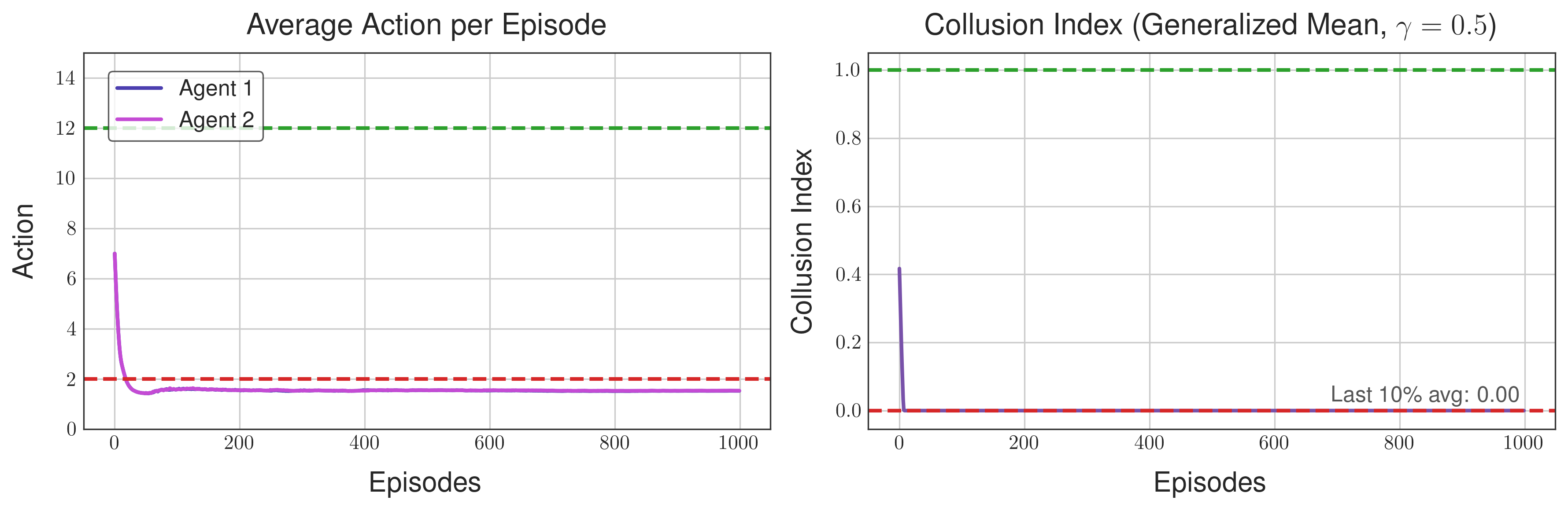}
        \caption{Training run evolution}
        \label{fig:compPPO:training}
    \end{subfigure}

    \begin{subfigure}[h]{\linewidth}
        \centering
        \includegraphics[width=0.6\linewidth]{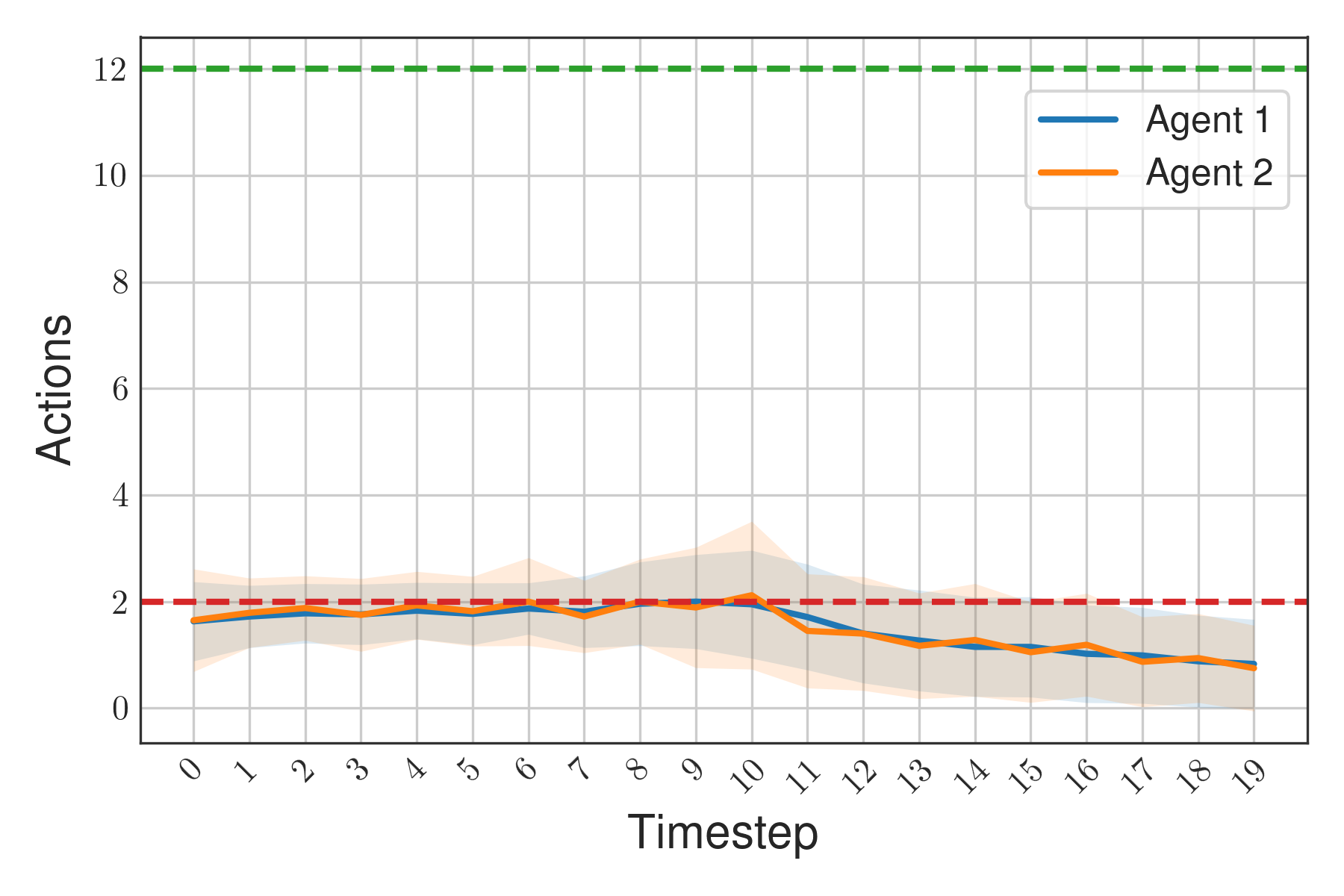}
        \caption{Evolution within episode}
        \label{fig:compPPO:episode}
    \end{subfigure}
    \caption{Competitive PPO: The evolution of a training run (a) and, once trained, within an episode (b) of two PPO learners trained with a very high ``number of environments'' parameter. They quickly converge to competition, and behave competitively throughout the entire episode.}
    \label{fig:compPPO}
   
\end{figure}

By increasing the noise in agents' learning targets, they can only learn best-response strategies, driving convergence toward competition. This is achieved by setting the ``number of environments'' parameter very high. PPO trains on a batch of data gathered from playing one or more parallel episodes against the same opponent, but with different random seeds. While our model has deterministic transitions, PPO has a stochastic policy, such that these episodes have different evolutions. With many different episodes in the learning batch, the PPO agents are likely not able to discern the opponent's underlying policy well and adapt to it, instead learning the (Nash) best response of competition. \Cref{fig:compPPO:training} shows the quick initial convergence to competition that is then never deviated from. \Cref{fig:compPPO:episode} shows an evaluation episode of the trained learners, who play (imperfect) competition throughout the episode. 

\subsection{Unconstrained DQN learners}\label{sec:appendix_unconstrained_DQN}
\begin{figure}[htbp]
    \centering
    \begin{subfigure}[t]{\linewidth}
        \centering
        \includegraphics[width=\linewidth]{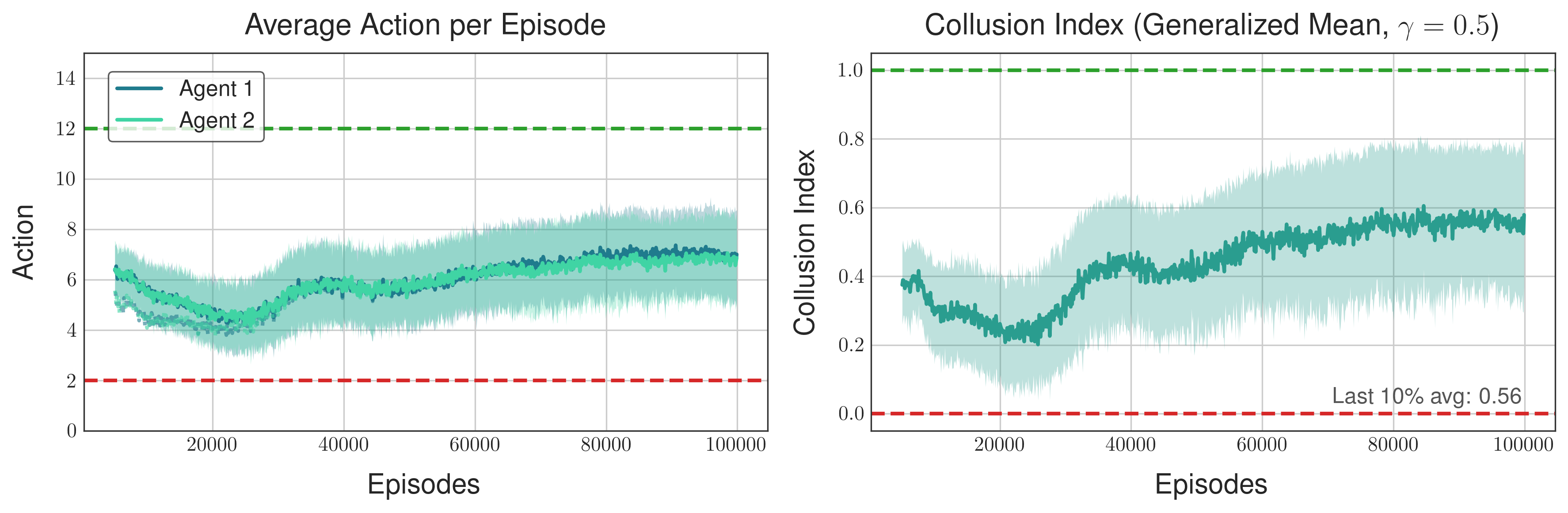}
        \caption{Training run evolution}
        \label{fig:unconst_DQN:collusion}
    \end{subfigure}

    \centering
    \begin{subfigure}[b]{\linewidth}
        \centering
        \includegraphics[width=\linewidth]{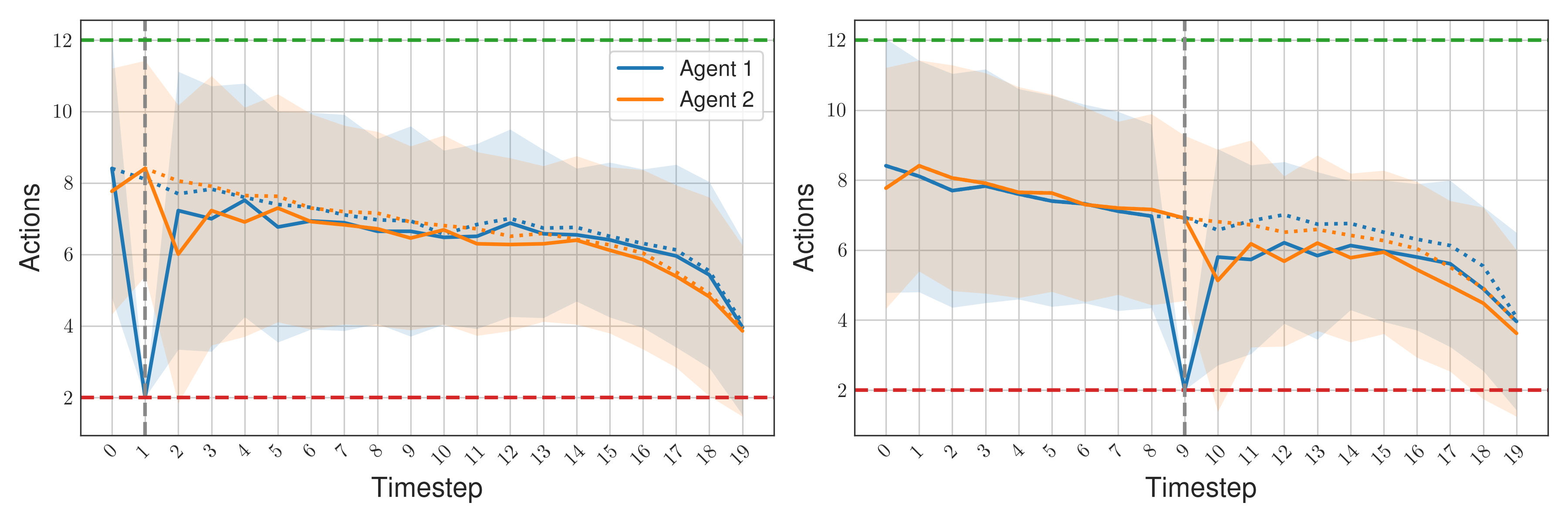}
        \caption{Forced deviation within episode}
        \label{fig:unconst_DQN:forced_deviation}
    \end{subfigure}
    \caption{Unconstrained inventory: We train two DQN learners with an unconstrained inventory and plot the evolution of a training run (a) and the dynamics within an episode where one agent is forced to deviate to competition at time $t=1$ and $t=9$ (b). We observe stronger collusion than in the constrained case, and larger reactions to forced deviations.}
    \label{fig:unconst_DQN}
\end{figure}

We analyze the setup of two DQN learners in an environment where inventory constraints are \emph{not} simulated. \Cref{fig:unconst_DQN:collusion} shows the evolution of the training run. The learners show stronger collusion and keep increasing collusion throughout the entire run. This is different to the constrained setting, where collusion stops increasing once a stable level is hit (even if training time is extended).

\Cref{fig:unconst_DQN:forced_deviation} shows the behavior of the trained agents during an evaluation episode. Unlike in the constrained setting, where collusion was built intra-episode, the unconstrained learners start the episode by already behaving collusively. Their response to forced deviation is stronger than in the constrained case, reacting more competitively. They display the same backward induction-type behavior, having learned that deviation toward the end of the episode is unable to be punished and therefore less risky.

\subsection{Impact of agent deviation on episode profits}
\label{sec:appendix_deviation_analysis}
\begin{figure}[t]
    \centering
    \begin{subfigure}[t]{\linewidth}
        \centering
        \includegraphics[width=\linewidth]{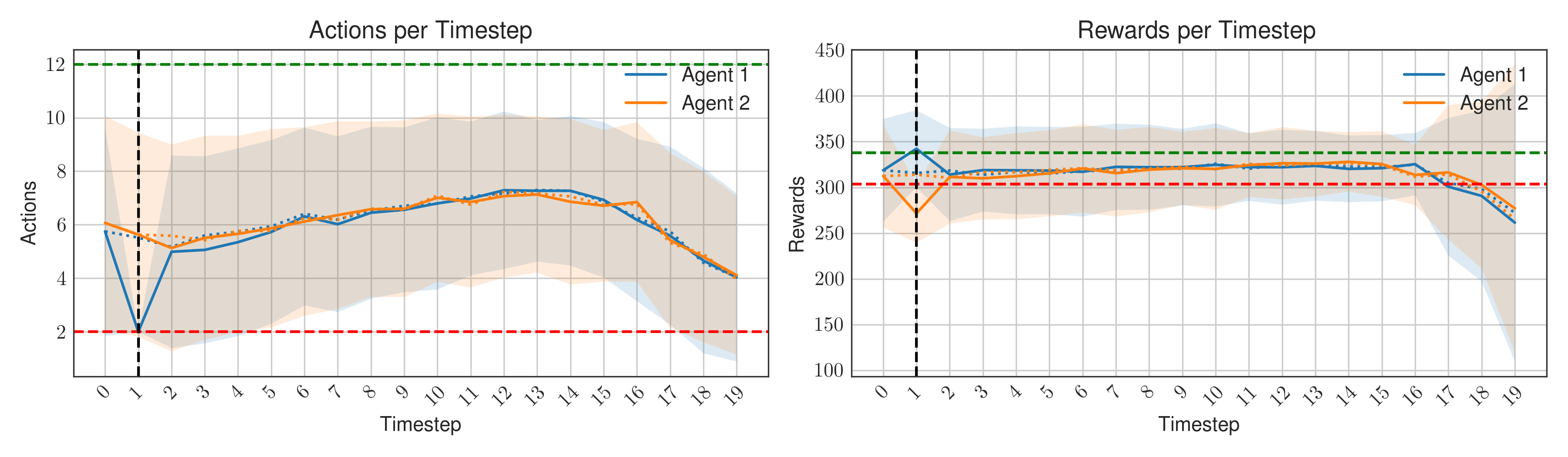}
        \vspace{0.5cm}
        \includegraphics[width=\linewidth]{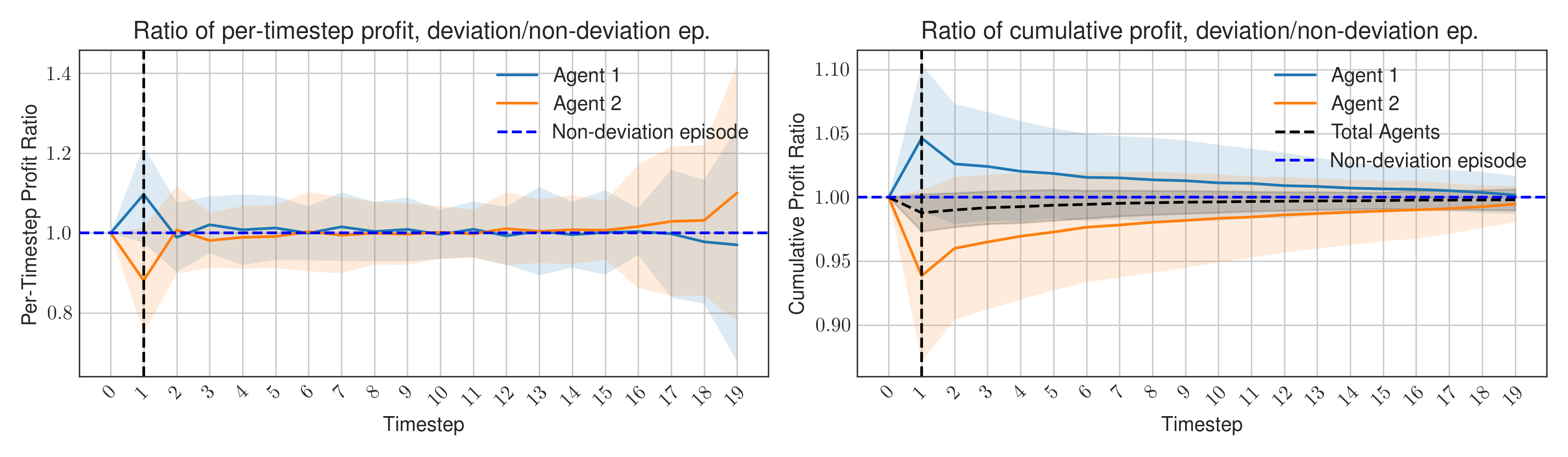}    
        \caption{Deviation at time t=1. End-of-episode total profit vs non-deviation: 99.81\%}
        \label{fig:DQN_full_forced_deviation:t1}
    \end{subfigure}
    \begin{subfigure}[t]{\linewidth}
        \centering
        \includegraphics[width=\linewidth]{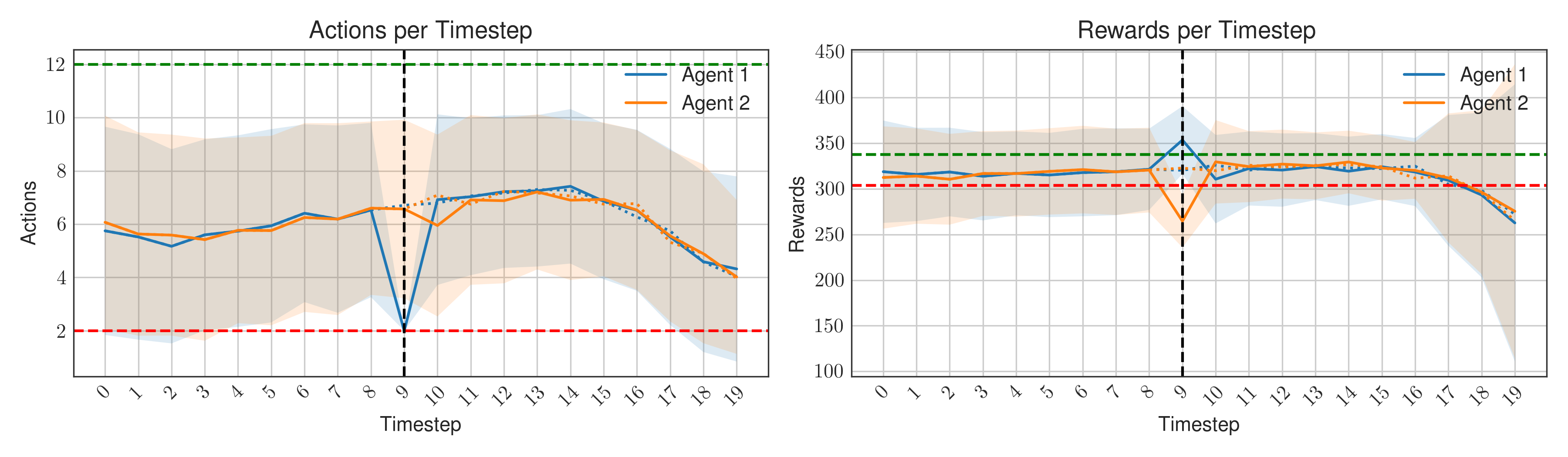}
        \vspace{0.5cm}
        \includegraphics[width=\linewidth]{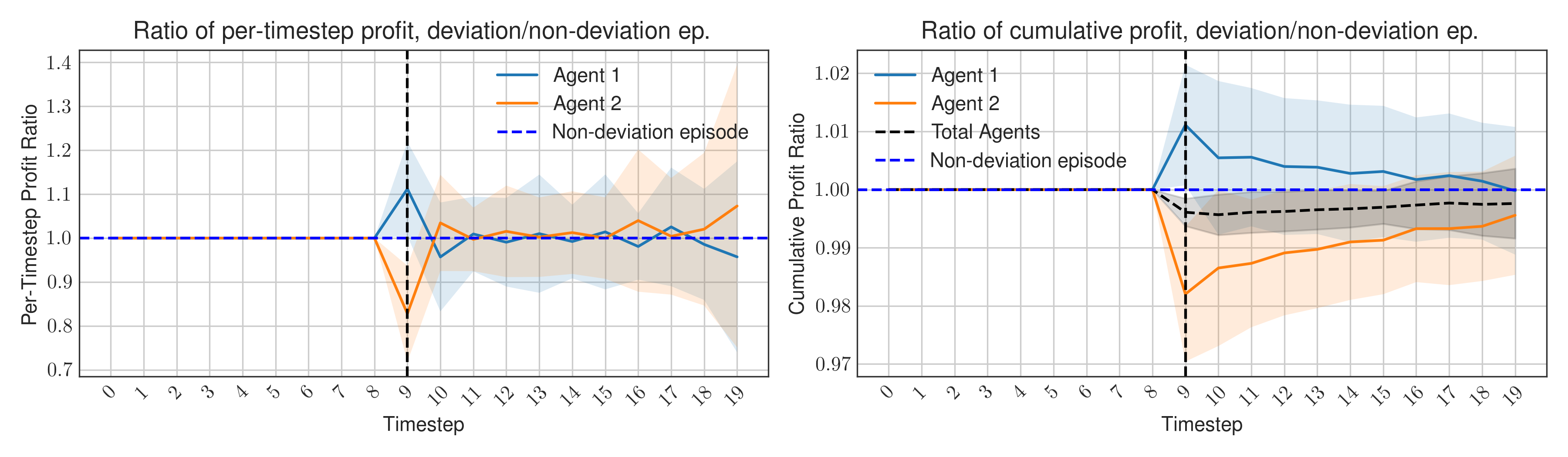}   
        \caption{Deviation at time t=9. End-of-episode total profit vs non-deviation: 99.76\%}
        \label{fig:DQN_full_forced_deviation:t9}
    \end{subfigure}
    \caption{``Profit impact of deviation''. Two trained DQN agents play an evaluation episode, with one agent being forced to deviate at different timesteps. We show the effect on individual and cumulative agent profit per single timestep, and over the entire evaluation episode. Total episode profit remains virtually unchanged by single deviations.}
    \label{fig:DQN_full_forced_deviation}
\end{figure}

\Cref{fig:DQN_full_forced_deviation} shows the evolution of two DQN agents' profits when forcing one agent to deviate. We observe that the deviating agent's temporary profit is tempered by the punishment in response, with both agents quickly returning to their normal strategy. The end-of-episode total profit is merely $0.2\%$ lower than without the deviation, with the deviating agent only breaking even and the non-deviating agent taking a slight loss.

\subsection{Additional hyperparameter comparisons}\label{sec:appendix_more_hparams}
\begin{figure}%

    \begin{subfigure}[h]{\linewidth}
        \centering
        \includegraphics[width=\linewidth]{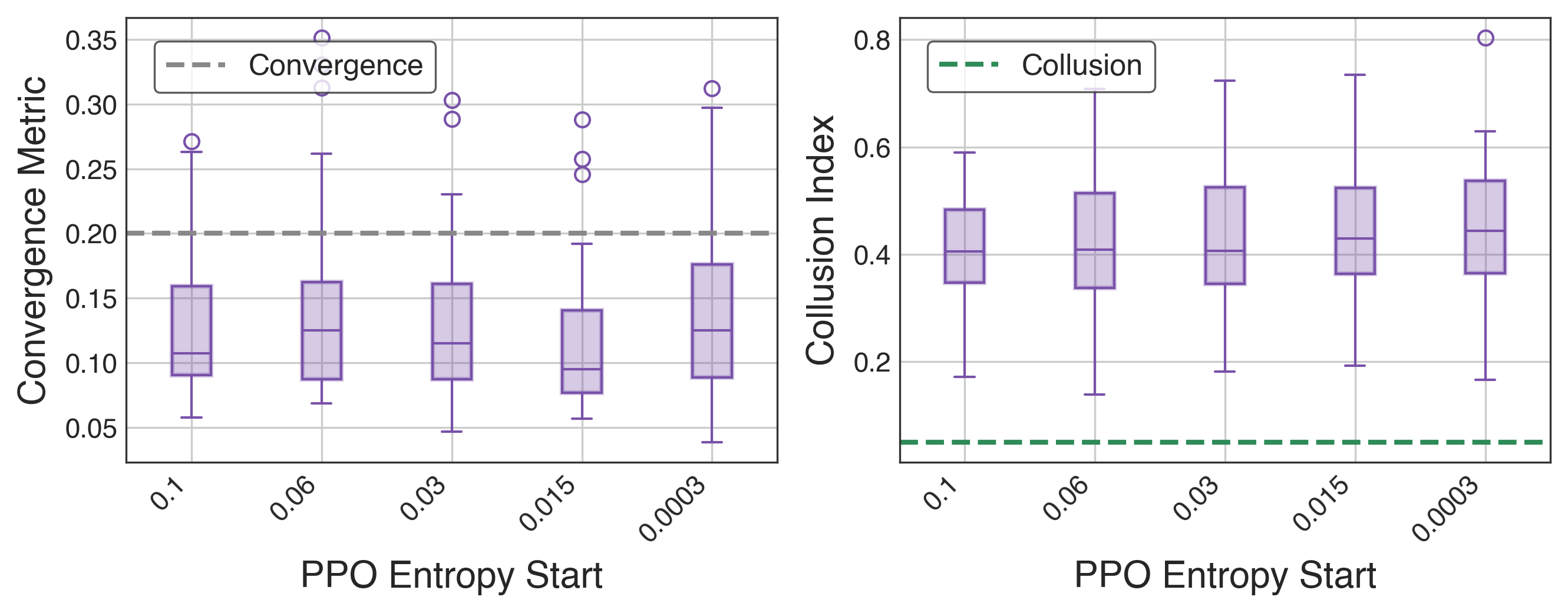}
        \label{fig:PPO_entropy_boxplot}
    \end{subfigure}
    
    \begin{subfigure}[h]{\linewidth}
    
        \centering
        \includegraphics[width=\linewidth]{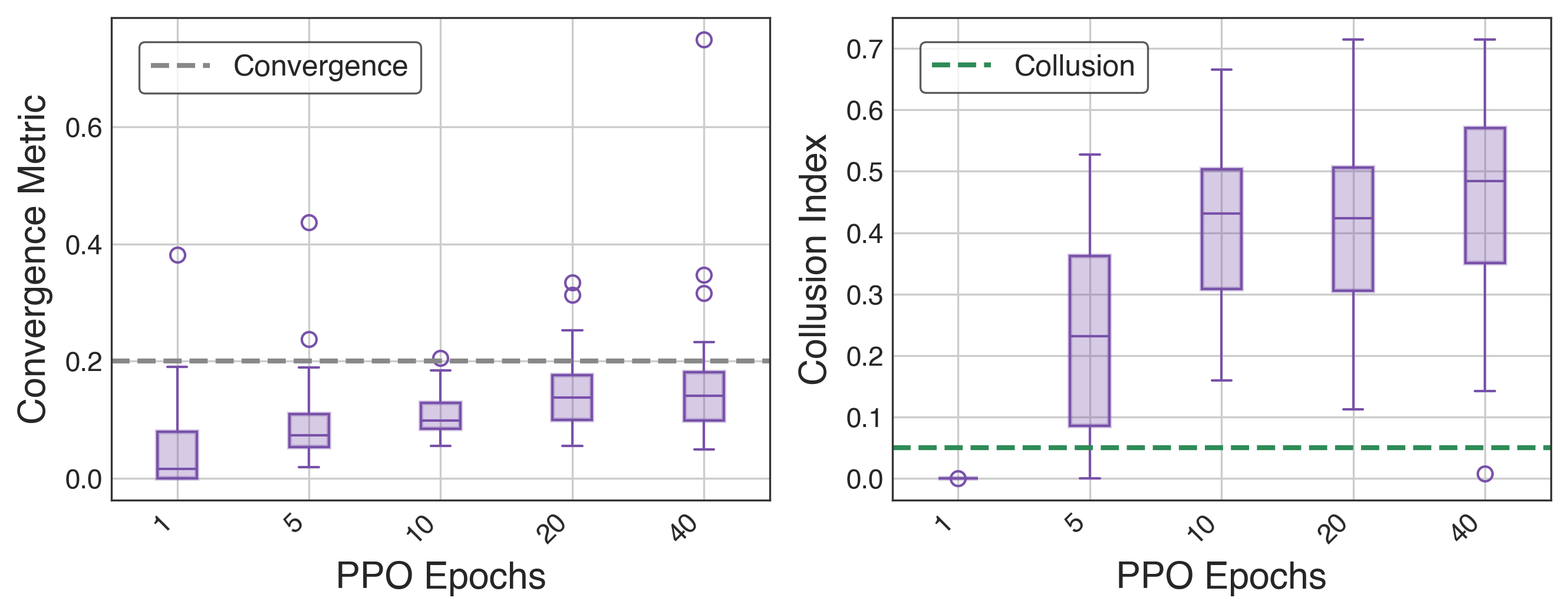}
        \label{fig:PPO_epochs_boxplot}
    \end{subfigure}
    
    \begin{subfigure}[h]{\linewidth}
    
        \centering
        \includegraphics[width=\linewidth]{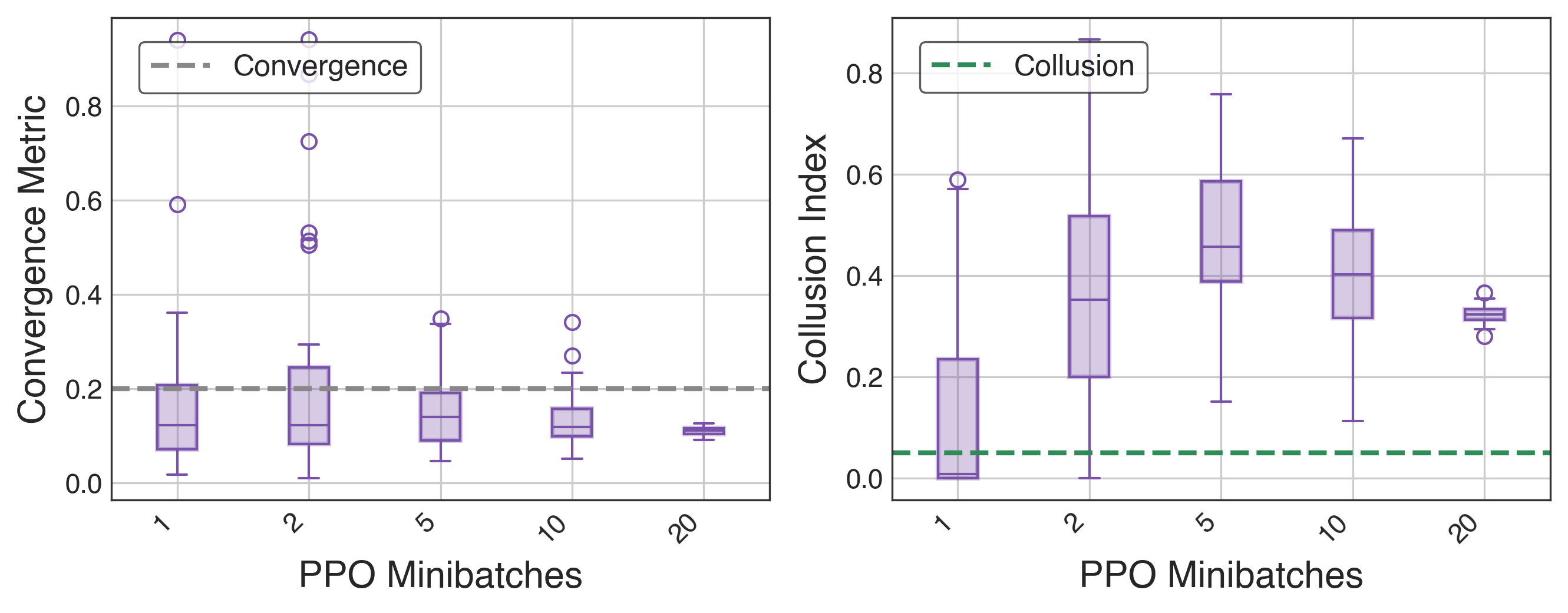}
        \label{fig:PPO_minibatches_boxplot}
    \end{subfigure}
    \caption{Convergence and collusion metrics for PPO training runs with varied starting entropy coefficient (top), number of epochs per training step (middle) and number of minibatches per training step (bottom). Collusion is robust against starting entropy and increases with more epochs or minibatches per training step.}
    \label{fig:appdx_PPO_boxplots}

\end{figure}

\begin{figure}%
    \begin{subfigure}[h]{\linewidth}
        \centering
        \includegraphics[width=\linewidth]{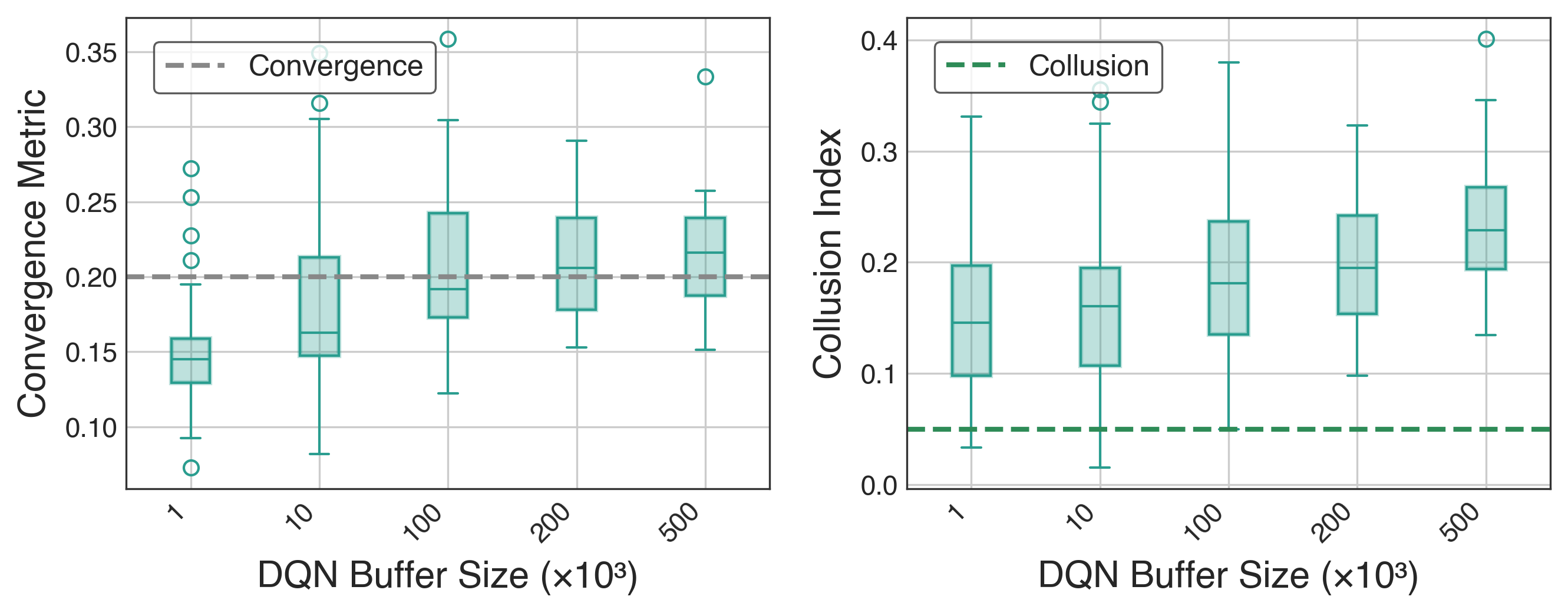}
        \label{fig:DQN_buffer_size_boxplot}
    \end{subfigure}
    
    \begin{subfigure}[h]{\linewidth}
        \centering
        \includegraphics[width=\linewidth]{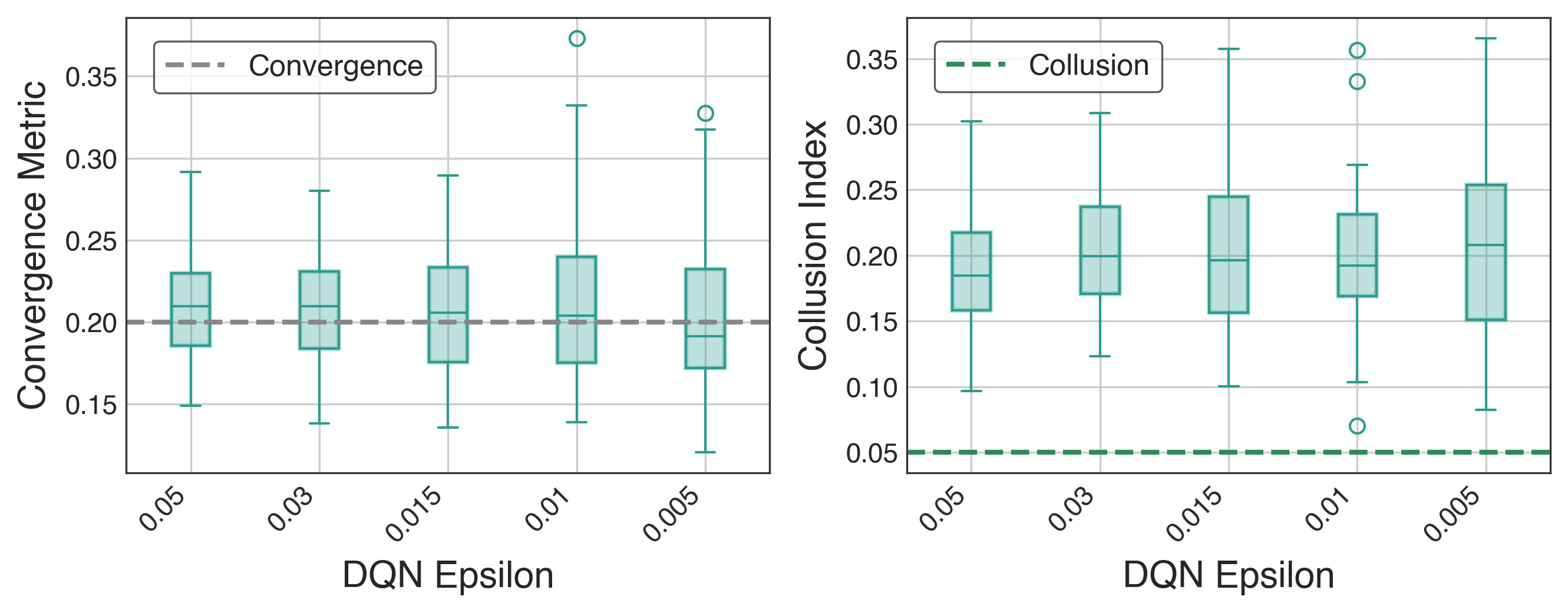}
        \label{fig:DQN_epsilon_boxplot}
    \end{subfigure}
    
    \begin{subfigure}[h]{\linewidth}
        \centering
        \includegraphics[width=\linewidth]{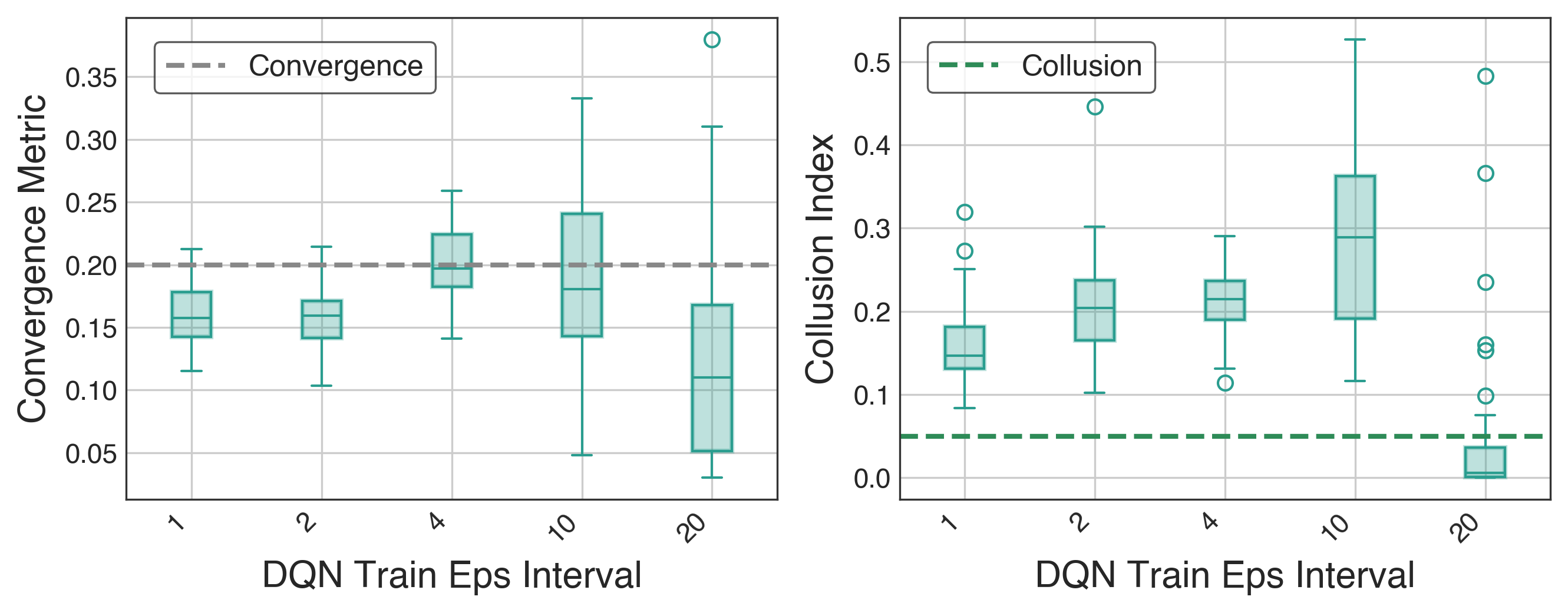}
        \label{fig:DQN_train_eps_interval_boxplot}
    \end{subfigure}
    \caption{Convergence and collusion metrics for DQN training runs with varied buffer size in thousands (top), exploration epsilon (middle) and length of interval between training episodes (bottom). Larger buffer sizes increase collusion but reduce convergence, lower epsilon slightly worsens convergence without affecting collusion, and longer intervals improve convergence and collusion up to a point before becoming too sparse for learning.}
    \label{fig:appdx_DQN_boxplots}
\end{figure}

We show some additional plots of agent hyperparameter behavior in \Cref{fig:appdx_PPO_boxplots} for PPO and \Cref{fig:appdx_DQN_boxplots} for DQN.

Beginning with PPO, scaling the amount of initial entropy has a marginal effect on both convergence and collusiveness. The number of epochs, on the other hand, has a much bigger impact. More epochs of training per training step on the same batch of data allows PPO to fit their strategy to their opponent's much more effectively, increasing collusiveness while slightly reducing convergence. Lastly, increasing the number of minibatches and thus frequency of gradient updates helps PPO converge, but it does hurt collusiveness, which could be explained from the increased noise from smaller batches. A very low number of minibatches sees very stable training -- perhaps too stable to effectively explore collusion.

DQN behaves similarly with regards to exploration and stable targets. A larger buffer size reduces convergence due to the increased variance in experiences that can be sampled (which are less up-to-date as buffer size increases). Larger buffers do help with establishing collusion, though, perhaps precisely because singular opponent deviations are less likely to be included in the next gradient step. We further observe that there is no strong dependence on initial exploration epsilon for either convergence or collusion. On the other hand, the interval between training episodes (as opposed to episodes where experience is gathered without a gradient update, i.e. only filling the replay buffer) does matter. Decreasing training frequency, and thus likelihood of immediate response to an opponent deviation increases collusive tendencies, but there is a limit -- training too infrequently increases instability in the targets again and leaves DQN unable to react to positive exploratory moves.

\clearpage

\clearpage


\end{document}

%% file: header.tex
\usepackage{authblk}
\usepackage{graphicx}
\usepackage{subcaption}
\usepackage[T1]{fontenc}    

\usepackage[hidelinks]{hyperref}       
\usepackage{booktabs}       
\usepackage{nicefrac}       
\usepackage[final,nopatch=footnote]{microtype}      
\usepackage{xcolor}         
\usepackage{algorithm}
\usepackage{algorithmic}

\usepackage[hang,flushmargin]{footmisc}

\graphicspath{arxiv/}

\usepackage[english]{babel}
\usepackage[utf8]{inputenc}
\usepackage{parskip}
\usepackage{comment}       
\usepackage{setspace}
\usepackage{csquotes}

\usepackage[backend=biber,natbib=true,style=apa,bibstyle=authoryear]{biblatex}
\DeclareFieldFormat[misc]{title}{\mkbibquote{#1}}
\usepackage{amsmath,amssymb,amsfonts,mathrsfs,bbm}
\usepackage{amsthm}
\usepackage{thmtools}
\usepackage{thm-restate}
\usepackage{bm}

\usepackage[capitalize]{cleveref} %
\usepackage{xr}
\usepackage{url}

\usepackage{enumitem}
\usepackage{mathtools}
\usepackage[h]{esvect}
\usepackage{array}
\usepackage{minitoc}

\usepackage{array}

\usepackage{subcaption}
\usepackage{siunitx}    %

\theoremstyle{plain}
\newtheorem{theorem}{Theorem}[section]

\newtheorem{lemma}[theorem]{Lemma}

\theoremstyle{definition}
\newtheorem{definition}[theorem]{Definition}

\theoremstyle{remark}

\DeclareMathOperator{\sgn}{sgn}

\newcommand{\NN}{\mathbb{N}}

\newcommand{\EE}{\mathbb{E}}
\newcommand{\numepisodes}{E}
\newcommand{\statespace}{\mathcal{S}}
\newcommand{\agentstate}{s}

\newcommand{\actionspace}{\mathcal{A}}

\newcommand{\action}{a}
\newcommand{\transitionfunc}{P}
\newcommand{\rewardfunc}{R}
\newcommand{\reward}{R}

\newcommand{\policy}{\pi}

\newcommand{\agentset}{N}
\newcommand{\numagents}{n}

\newcommand{\timehorizon}{T}

\newcommand{\profitgain}{\Delta}
\newcommand{\collindex}{\Delta}
\newcommand{\quality}{\alpha}
\newcommand{\horizontaldiff}{\mu}
\newcommand{\cost}{c}
\newcommand{\price}{p}
\newcommand{\demand}{d}
\newcommand{\quantity}{q}
\newcommand{\inventory}{x}
\newcommand{\capacity}{I}
\newcommand{\demandscalingfactor}{\lambda}